\documentclass[nofootinbib,twocolumn,showpacs,amsmath,amstex,amssymb,mathfonts,superscriptaddress,prb]{revtex4-1}

\usepackage{graphicx}
\usepackage{color}
\usepackage{bm}
\usepackage{amsmath}
\usepackage{amssymb}
\usepackage{amsthm}
\usepackage{amsfonts}
\usepackage[normalem]{ulem}
\usepackage[colorlinks,bookmarks=true,citecolor=blue,linkcolor=red,urlcolor=blue]{hyperref}

\usepackage{bbm}

\newcommand*{\id}{{\normalfont\hbox{1\kern-0.15em \vrule width .8pt depth-.5pt}}}

\newcommand{\be}{\begin{equation}}
\newcommand{\ee}{\end{equation}}
\newcommand{\bq}{\begin{eqnarray}}
\newcommand{\eq}{\end{eqnarray}}

\newcommand{\rf}[1]{(\ref{#1})}

\newcommand{\p}{{\boldsymbol p}}
\newcommand{\q}{{\boldsymbol q}}
\newcommand{\bs}[1]{\boldsymbol{#1}}

\newcommand{\ket}[1]{\left |#1 \right\rangle}
\newcommand{\bra}[1]{\left \langle #1 \right |}

\theoremstyle{theorem}

\theoremstyle{theorem}

\theoremstyle{definition}

\theoremstyle{definition}

\theoremstyle{remark}

\theoremstyle{theorem}

\makeatletter
\def\@fnsymbol#1{\ensuremath{\ifcase#1\or \dagger\or \ddagger\or
   \mathsection\or \mathparagraph\or \|\or **\or \dagger\dagger
   \or \ddagger\ddagger \else\@ctrerr\fi}}
    \makeatother

\begin{document}
\title{Geometric description of the Kitaev honeycomb lattice model}

\author{Ashk Farjami}
\thanks{A.F. and M.D.H. contributed equally to this work}
\affiliation{School of Physics and Astronomy, University of Leeds, Leeds, LS2 9JT, United Kingdom}

\author{Matthew D. Horner}
\thanks{A.F. and M.D.H. contributed equally to this work}
\affiliation{School of Physics and Astronomy, University of Leeds, Leeds, LS2 9JT, United Kingdom}

\author{Chris N. Self}
\affiliation{School of Physics and Astronomy, University of Leeds, Leeds, LS2 9JT, United Kingdom}
\affiliation{Blackett Laboratory, Imperial College London, Prince Consort Road, London, SW7 2AZ, United Kingdom}

\author{Zlatko Papi\'c}
\affiliation{School of Physics and Astronomy, University of Leeds, Leeds, LS2 9JT, United Kingdom}

\author{Jiannis K. Pachos}
\affiliation{School of Physics and Astronomy, University of Leeds, Leeds, LS2 9JT, United Kingdom}

\date{\today }

\begin{abstract}

It is widely accepted that topological superconductors can only have an effective interpretation in terms of curved geometry rather than gauge fields due to their charge neutrality. This approach is commonly employed in order to investigate their properties, such as the behaviour of their energy currents. Nevertheless, it is not known how accurately curved geometry can describe actual microscopic models. Here, we demonstrate that the low-energy properties of the Kitaev honeycomb lattice model---a topological superconductor that supports localised Majorana zero modes at its vortex excitations---are faithfully described in terms of Riemann-Cartan geometry. In particular, we show analytically that the continuum limit of the model is given in terms of the Majorana version of the Dirac Hamiltonian coupled to both curvature and torsion. We numerically establish the accuracy of the geometric description for a wide variety of couplings of the microscopic model. Our work opens up the opportunity to accurately predict dynamical properties of the Kitaev model from its effective geometric description.

\end{abstract}

\maketitle


\section{Introduction}

In recent years there has been a surge of interest in the geometrical degrees of freedom that characterise the response of topologically-ordered phases of matter beyond the well-known limit governed by an effective topological quantum field theory. An important class of such systems, the fractional quantum Hall (FQH) states, have been understood to exhibit a universal response to the variations of ambient geometry, leading to many fruitful investigations of an interplay between topology and geometry in these strongly-correlated systems
~\cite{wen1992shift, avron1995viscosity, read2009non, HaldaneViscosity, abanov2014electromagnetic, gromov2015framing, BradlynRead, CanLaskinWiegmann, klevtsov2015geometric, hughes2011torsional, gromov2014density}. In particular, the neutral collective mode of FQH systems~\cite{GMP85} has been described as a fluctuating spacetime metric~\cite{HaldaneGeometry, GromovSon,wiegmann2017nonlinear}. 
On the other hand, a recent study~\cite{Golan} has shown that by minimally coupling a spinless $p$-wave superconductor on a square lattice to an electromagnetic field, the continuum limit takes the form of a Dirac Hamiltonian defined on a spacetime with both curvature and torsion. Such curved spaces with torsion are called \textit{Riemann-Cartan} spacetimes~\cite{Hehl}. Riemann-Cartan geometry also naturally arises in the theory of defects in lattices, whereby disclinations and dislocations in the continuum limit are described by curvature and torsion, respectively~\cite{Katanaev, deJuan}, which has been investigated in strained graphene~\cite{Wagner,deJuan,deJuan2}. Other techniques from quantum gravity have also been employed in condensed matter such as the holographic correspondence or AdS/CFT correspondence. Recently, the holographic correspondence has been used to model gapless modes living on the defect lines of class D topological superconductors\cite{Palumbo} and to determine the specific heat of a two-dimensional interacting gapless Majorana system\cite{Maraner}. Moreover, the emergence of gravitational anomalies has been considered in topological superconductors~ \cite{Jaakko}. Building upon Luttinger's proposal~\cite{Luttinger}, gravitational techniques applied to the thermal Hall effect have also attracted many theoretical\cite{Nakai, Cooper, ReadGreen, Wang2, Ryu, Qin, Shitade} and experimental~\cite{Jezouin,Banerjee2} investigations.

Nevertheless, despite much progress in the investigation of geometric effects in the continuum field theory description, the study of Riemann-Cartan geometry in microscopic, solvable lattice models has received less attention. In this paper, we investigate geometric description of the Kitaev honeycomb lattice model (KHLM)~\cite{Kitaev}, the well-known two-dimensional (2D) model of interacting spin-$\frac{1}{2}$ particles that gives rise to a quantum spin liquid phase with topological order. A salient feature of the KHLM is that it can support non-Abelian anyons in the form of Majorana zero modes (MZMs) trapped at its vortices~\cite{Kitaev,Kitaev2,Otten,Vidal3,Vidal4}. 
Similar to the FQH effect~\cite{Moore}, the KHLM is both topologically ordered in the sense that it can support anyonic excitations and it is a topological phase categorised by a non-trivial Chern number~\cite{Kitaev,Vidal2}. Unlike the FQH effect, the KHLM is exactly solvable, which has provided unique opportunities to analytically probe its anyonic properties~\cite{Ville1,Vidal3, Vidal4}, its topological edge currents~\cite{Chris1}, its finite temperature behaviour~\cite{Chris2,Ville3,Ville4,Nasu,Nasu2,Nasu3} and to investigate dimer limits of the model~ \cite{Vidal,Vidal5}. Moreover, many features of the KHLM are recognised in experimentally realisable materials, such as complex iridium oxides \cite{Chaloupka, Choi, Jackeli} or ruthenium chloride \cite{Banerjee}. This makes the KHLM of interest to numerous theoretical and experimental investigations.

In this paper we address the following question: can we use the KHLM to simulate Majorana fermions embedded in a Riemann-Cartan spacetime? To answer this question, we allow the couplings of the KHLM to take general configurations that are anisotropic and inhomogeneous, while leaving the lattice configuration of the model unaffected. We demonstrate that in this case the low energy limit of the model can be effectively described by massless Majorana spinors obeying the Dirac equation embedded in a Riemann-Cartan spacetime which is locally Lorentz invariant. Moreover, the Majorana spinors are coupled to a non-trivial torsion. It is important to stress that this geometry emerges purely from distortions in the couplings of the system and {\em not} from the geometry of the lattice itself.
 
We shall see that choosing the two-spin couplings to be anisotropic gives rise to a general two-dimensional metric. When these couplings are varying in space they can give rise to an arbitrary curved space. Moreover, the three-spin interactions generate the torsion of the Riemann-Cartan geometry. As a result, the KHLM can be effectively described by massless relativistic Majorana spinors. These particles become massive when a Kekul\'e distortion is introduced in the two-spin couplings~\cite{Yang}. We numerically investigate the phase diagram produced by the torsion and mass terms that support competing topological phases.

To quantify how faithful the geometric description is to the KHLM we consider two physical quantities of the original microscopic model. First, we investigate the spatial distribution of the quantum correlations in the ground state of the model. Second, we analyse the shape of a Majorana zero mode bounded by a vortex. While we vary the couplings of the model away from the isotropic regime we numerically observe that the geometric distortion of the quantum correlations as well as shape of the zero mode follow the theoretically predicted geometric configurations. This agreement is faithful through  most of the non-Abelian phase only breaking down near the phase transition boundaries. Hence, the Riemann-Cartan description can be employed to accurately describe the behaviour of the KHLM in a quantum field theory language. This opens up the exciting possibility to theoretically study response properties of KHLM, such as the energy currents as a function of coupling distortions or temperature gradients, in a quantitative way.

The paper is organised as follows. In Section \ref{KHLM} we provide a self-contained introduction to the KHLM and demonstrate that the low energy limit of the model is described by massless Majorana fermions satisfying the Dirac Hamiltonian. In Section \ref{Riemann-Cartan} we review the relevant theory of Majorana spinors defined on a Riemann-Cartan geometry and derive the corresponding Hamiltonian. In Section \ref{Riemann-Cartan KHLM} we demonstrate that the low energy limit of the KHLM can be faithfully described by massless Majorana fermions propagating on a Riemann-Cartan background, with geometric characteristics fully determined by the coupling constants of the model. We also present the Kekul\'e distortion modification to the KHLM, which is responsible for generating mass in the continuum limit. In Section \ref{numerics} we present numerical results of correlation functions and zero mode profiles for various coupling configurations of the KHLM in order to verify its non-trivial description in terms of a metric. Finally in Section \ref{sec:Conclusions} we present the conclusions and an outlook of our work.


\section{Kitaev honeycomb lattice model in the continuum \label{KHLM}}

Similar to graphene\cite{Neto,DiVincenzo,Semenoff}, the KHLM with isotropic, homogeneous couplings has a continuum limit that is given in terms of a Dirac Hamiltonian. To reveal this feature we consider the low energy sector of the model where long wavelengths are dominant and the lattice spacing is negligible. In this low energy regime the KHLM can be efficiently described by a linear energy dispersion relation. This linearity between energy and momentum can be modelled by the Dirac Hamiltonian. For the case of the KHLM, this Dirac Hamiltonian can take a purely imaginary form signalling that the effective degrees of freedom of the model are Majorana fermions. As a result, the model is faithfully described by a quantum field theory of relativistic Majorana fermions. We will review this derivation here in order to set our notation for the later sections.

\begin{figure}[t]
\center
\includegraphics[width =\linewidth]{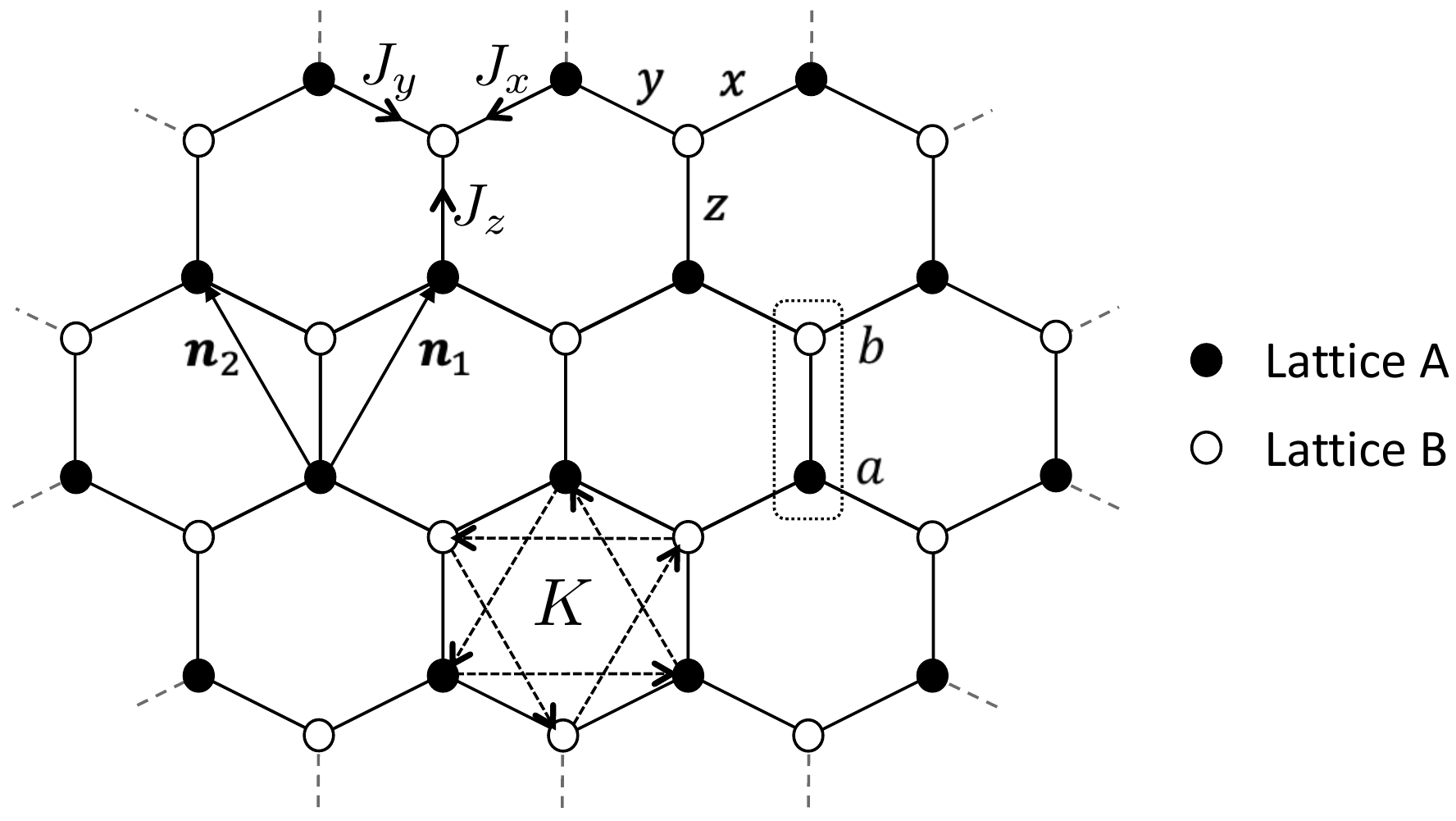}
\caption{The honeycomb lattice with Majorana fermions tunnelling between nearest neighbouring sites with couplings $J_x$, $J_y$ and $J_z$ depending on the direction of the link. Tunnelling between next-to-nearest neighbouring sites with coupling $K$ is also indicated. The honeycomb lattice comprises two triangular sub-lattices, $A$ and $B$, denoted by full and empty circles, respectively. We take the unit cell along the $z$-links. The translation vectors between sites of the same sub-lattices are $\boldsymbol{n}_1=(\frac{\sqrt3}{2},\frac{3}{2})$ and $\boldsymbol{n}_2=(-\frac{\sqrt3}{2},\frac{3}{2})$. The orientations of the nearest tunnelings (from $A$ to $B$ sites) and next-to-nearest tunnelings (anticlockwise) are indicated.}
\label{fig:honeycomb}
\end{figure}

Kitaev's honeycomb lattice is an exactly solvable model that describes interacting spin-$\frac{1}{2}$ particles residing on the vertices of a honeycomb lattice~\cite{Kitaev}. The spin couplings consist of nearest and next-to-nearest neighbour interactions described by the Hamiltonian
\begin{equation}
\begin{aligned}
H  = & -4\bigg(J_x \sum_\text{$x$ links} \sigma^x_i \sigma^x_j + J_y \sum_\text{$y$ links} \sigma^y_i \sigma^y_j + J_z \sum_\text{$z$ links} \sigma^z_i \sigma^z_j \\
& + K \sum_{(i,j,k)} \sigma^x_i \sigma^y_j \sigma^z_k\bigg),
\label{eq:Kitaev spin}
\end{aligned}
\end{equation}
where the orientation of the $x$-links, $y$-links and $z$-links are indicated in Fig.~\ref{fig:honeycomb}. In Eq.~\rf{eq:Kitaev spin} we have introduced a factor of 4 compared to Ref.~\onlinecite{Kitaev} in order to simplify the subsequent algebra. The three spin interaction acts between three successive spins around a hexagonal plaquette~\cite{Kitaev}. 
 For certain values of the couplings $\{J_i \}$ and $K$, 
the ground state of the model is a quantum spin liquid~\cite{Savary, Balents, Knolle}. This state exhibits exotic quantum order due to long-range entanglement, which leads to spin-fractionalised excitations that behave as Abelian or non-Abelian anyons~\cite{Kitaev}.

By employing an appropriate fermionisation procedure, the spin Hamiltonian can be brought to a form describing Majorana fermions coupled to a $\mathbb{Z}_2$ gauge field. This gauge field can encode vortices which are eigenstates of the original spin Hamiltonian. Here, we shall restrict ourselves to the {\em no-vortex sector}, where the gauge field has a trivial configuration. In this case, the Hamiltonian takes the form
\begin{equation}
H = {i}\left( \sum_{\langle i,j\rangle}  2J_{ij}c_i c_j + 2K \sum_{\langle\langle i,j\rangle\rangle} c_i c_j \right), 
\label{eq:Kitaev}
\end{equation}
where $\{ c_i \}$ are Majorana fermions sitting on the site $i$.
The first term of the Hamiltonian is a sum over nearest neighbours whilst the second term is a sum over next-to-nearest neighbours. The ordering of the Majorana operators for the nearest and next-to-nearest couplings is as shown in Fig.~\ref{fig:honeycomb}.

The honeycomb lattice contains two triangular sub-lattices, $A$ and $B$, denoted in Fig.~\ref{fig:honeycomb} by full and empty circles, respectively. With this identification, we denote the Majoranas of sub-lattice $A$ as $c^a_i$ and those of sub-lattice $B$ as $c_i^b$, where now the index $i$ labels a unit cell of two neighbouring sites along the $z$-direction, as shown in Fig.~\ref{fig:honeycomb}. The honeycomb lattice model is periodic with respect to this unit cell. We are able to diagonalise the Hamiltonian in Majorana form by taking its Fourier transform, through the definitions $c^{a/b}_i = \int \mathrm{d}^2q e^{-i{\bs q}\cdot {\bs r}_i}c^{a/b}_{\bs{q}}$, where ${\bs r}_i$ is the position of the cell $i$. If we define the two-component spinor $\psi_{\q} = (c^a_{\q} \ ic^b_{\q})^\mathrm{T}$, then the Hamiltonian takes the form $H = \int \mathrm{d}^2q \psi_\q^\dagger h(\q) \psi_\q $, with the single-particle Hamiltonian $h(\boldsymbol{q})$ given by
\be
h(\q) = 
\begin{pmatrix}
\Delta(\q) &-f(\q) \\
-f^*(\q) & - \Delta(\q) 
\end{pmatrix}, \label{eq:latticeham}
\ee
where 
\be
f(\q) = 2(J_xe^{i\q \cdot {\boldsymbol n}_1} + J_ye^{i \q \cdot  {\boldsymbol n}_2} + J_z),
\ee
and
\begin{equation}
\begin{aligned}
\Delta(\q) = 2K [&-\sin(\q \cdot  {\boldsymbol n}_1) + \sin(\q \cdot  {\boldsymbol n}_2) \\
&+ \sin(\q \cdot ( {\boldsymbol n}_1 -  {\boldsymbol n}_2) )].
\end{aligned}
\end{equation} 

\begin{figure}[t]
\center
\includegraphics[width =\linewidth]{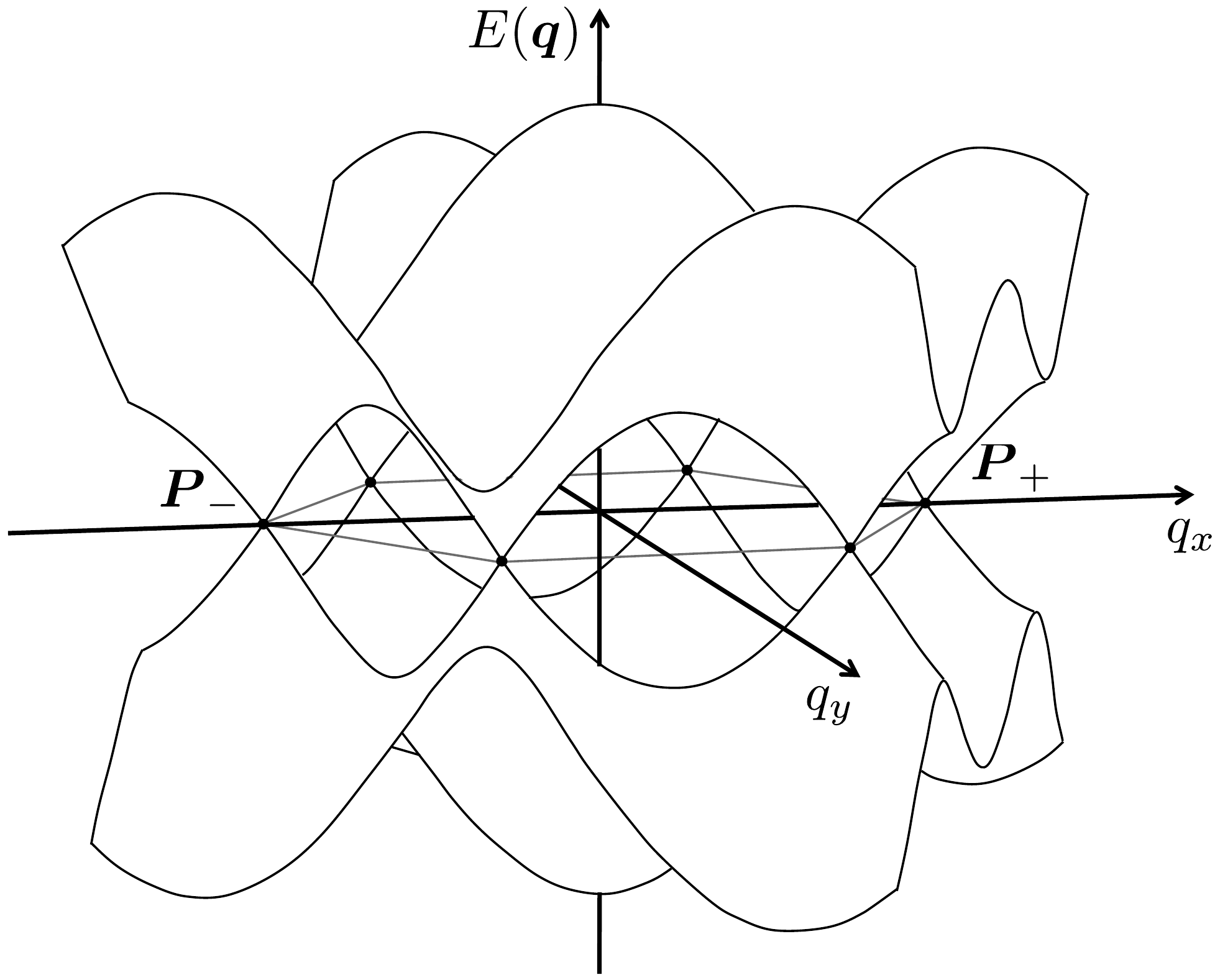}
\caption{The dispersion relation $E(\q)$ for the honeycomb Hamiltonian when $J_x=J_y=J_z=J$ and $K=0$. The Fermi points are the points where $E(\q) = 0$. The two inequivalent Fermi points within the Brillouin zone are given by $\boldsymbol{P}_+$ and $\boldsymbol{P}_-$.}
\label{fig:dispersion}
\end{figure}

For simplicity we initially consider the isotropic case where $J_x=J_y=J_z=J$. The dispersion relation of the single-particle Hamiltonian is given by
\begin{equation}
E(\q) = \pm \sqrt{\Delta^2(\q) + |f(\q)|^2}. 
\label{eq:dispersion}
\end{equation}
For $K=0$ the spectrum of $h(\q)$ is plotted in Fig.~\ref{fig:dispersion}. We see that the system has two independent Fermi points in the Brillouin zone located at
\begin{equation}
{\boldsymbol P}_\pm = \pm \left( \frac{4 \pi}{3\sqrt{3}} , 0 \right).
\end{equation} 
When $K\neq0$ this opens a gap in the dispersion of $2|\Delta(\boldsymbol{P}_\pm)|$, where $\Delta({\boldsymbol P}_\pm)=\pm\Delta$ and $\Delta=-3\sqrt{3}K$. The presence of Fermi points enables us to analyse the low energy properties of the Kitaev model and extract its behaviour in the continuum limit. 

In the regions close to the two Fermi points the Majorana fields have a linear dispersion relation much like the spectrum of graphene, as shown in Fig.~\ref{fig:dispersion}. Hence, at low enough energies the Hamiltonian $h(\q)$ can be effectively described by a Dirac Hamiltonian. In order to obtain this description, we substitute $\boldsymbol{q} = \boldsymbol{P}_\pm + \boldsymbol{p}$ for small $\boldsymbol{p}$ and Taylor expand $h(\q)$ around the Fermi points up to first order in $\boldsymbol{p}$. For convenience we define the Hamiltonian $h_\pm(\p) \equiv h({\boldsymbol P}_\pm + \p)$ for each Fermi point. For non-zero $K$ we obtain
\be
h_\pm(\p) = 3J ( \pm \sigma^x p_x - \sigma^y p_y )\mp 3\sqrt{3}K\sigma^z + O(p^2),
\ee
which act on the spinors $\psi_+(\p)=(c^a_+, ic^b_+)^\mathrm{T}$ and $\psi_-(\p)=(c^a_-, ic^b_-)^\mathrm{T} $, respectively, where $c^{a/b}_\pm(\boldsymbol{p})=c^{a/b}_{{\boldsymbol P}_\pm+\boldsymbol{p}}$. As these two Hamiltonians need to be considered simultaneously, we can treat the two Fermi points as different chiral degrees of freedom by defining the Dirac-like spinor $\Psi(\p) = (c^a_+ \ ic^b_+ \ ic^b_- \ c^a_-)^\mathrm{T}$. We take the direct sum of $h_+(\p)$ and $h_-(\p)$ in their respective bases defined by $\Psi$ to yield the total $4\times4$ Hamiltonian
\be
h_{\text{KHLM}}(\p) =3J (\sigma^z\otimes\sigma^x p_x - \sigma^z\otimes\sigma^y p_y) - 3\sqrt{3}K\mathbb{I}\otimes\sigma^z.
\label{eq:4by4H}
\ee
Note that we have rotated $h_-(\p)$ with a $\sigma^x$ rotation before combining it with $h_+(\p)$ to obtain $h_{\text{KHLM}}(\p)$.

The low energy limit given by \rf{eq:4by4H} suggests that we use the Dirac ${\bs \alpha}$ and $\beta$ matrices
\be
\boldsymbol{\alpha} = 
\begin{pmatrix}
\boldsymbol{\sigma} & 0 \\
0 & -\boldsymbol{\sigma}
\end{pmatrix}
= \sigma^z\otimes\boldsymbol{\sigma}, \quad
{\beta} = 
\begin{pmatrix}
0 &\mathbb{I} \\
\mathbb{I} & 0
\end{pmatrix}
= \sigma^x\otimes\mathbb{I},
\ee
where $\boldsymbol{\sigma} =(\sigma^x,\sigma^y,\sigma^z)$ are the Pauli matrices and $\mathbb{I}$ is the two-dimensional identity. The corresponding Dirac gamma matrices are defined by $\gamma^0 = \beta$ and $\boldsymbol{\gamma} = \beta^{-1} \boldsymbol{\alpha}$ where
\begin{equation}
\gamma^0 = \begin{pmatrix}
0 &\mathbb{I} \\
\mathbb{I} & 0
\end{pmatrix}
= \sigma^x\otimes\mathbb{I}, \quad
\boldsymbol{\gamma}=
\begin{pmatrix}
0 & -\boldsymbol{\sigma} \\
\boldsymbol{\sigma} & 0
\end{pmatrix}
= -i\sigma^y\otimes\boldsymbol{\sigma}. \label{eq:gamma_matrices}
\end{equation}
These matrices satisfy the Clifford algebra $\{ \gamma^a, \gamma^b \} = 2\eta^{ab}$, where Latin indices $a,b \in (0,1,2)$ and $\eta_{ab} = \mathrm{diag}(1,-1,-1)$ is the Minkowski metric. 

Using the gamma matrices, the Hamiltonian \rf{eq:4by4H} becomes
\begin{equation}
h_\text{KHLM}(\p)  = 3J \gamma^0(  \gamma^1 p_x -  \gamma^2 p_y ) -  3\sqrt{3}K i \gamma^1 \gamma^2 .
\label{eq:DiracKit}
\end{equation}
By its turn the many-body Hamiltonian of the low energy limit is given by
\be
H_\text{KHLM} =  \int \mathrm{d}^2p \; \Psi^\dagger(\boldsymbol{p}) \; h_\text{KHLM}(\p) \; \Psi(\boldsymbol{p}). 
\label{eq:totalH}
\ee
Hence, the low energy limit of the Kitaev model is given by the Dirac Hamiltonian with linear energy dispersion relation and an additional $K$ term that gives rise to an energy gap at the Fermi points.

To verify that the spinors of the Dirac operator are Majorana we first define charge conjugation. The charge conjugate of a Dirac spinor $\Psi$ is defined as $\Psi^{(c)}(\boldsymbol{p}) = C \Psi^*(-\boldsymbol{p})$, where $C$ is the unitary charge conjugation matrix  satisfying $C^\dagger \gamma^\mu C = - (\gamma^\mu)^*$ for all gamma matrices. In our chiral basis the charge conjugation matrix is given by $C = - \sigma^y \otimes \sigma^y$. With this and the fact that the Majorana modes obey $c_\pm^\dagger(\boldsymbol{p}) = c_{\mp}(- \boldsymbol{p})$, we see that the spinor $\Psi$ satisfies the neutrality condition $\Psi^{(c)}(\boldsymbol{p}) = \Psi(\boldsymbol{p})$ and is therefore a Majorana spinor. 



\section{Majorana Fields on $(2+1)$-dimensional Riemann-Cartan Geometry \label{Riemann-Cartan}}

As we have seen in the previous section, the continuum limit of the KHLM describes a Majorana field on a Minkowski spacetime. We now consider a Majorana field on a \textit{curved} spacetime. To proceed we first introduce the tools that allow us to transform from a flat space, where the usual Dirac Hamiltonian is defined, to a curved space. This can be achieved with the help of the dreibein and spin connection. In particular, we are interested in curved spacetimes that support torsion, i.e. deformations of spacetime that cannot be descried by a metric. These generalised spacetimes are called \textit{Riemann-Cartan} spacetimes. The following section closely follows references  \onlinecite{Nakahara} and \onlinecite{Carroll}.


\subsection{Riemann-Cartan Spacetime}

\subsubsection{Dreibein}
Consider a $(2+1)$-dimensional spacetime $M$ with coordinate system $(t,x,y)$. At every point on $M$ we have the coordinate basis vectors $\{ e_\mu = \partial_\mu \}$ of the tangent spaces, together with their corresponding dual basis vectors $\{ e^\mu = \mathrm{d}x^\mu \}$ satisfying $e^\mu(e_\nu) = \delta^\mu_\nu$. We use \textit{Greek} indices ranging over $t,x,y$ to represent components of tensors with respect to the coordinate basis. 
When discussing spinors on a general spacetime we need the notion of an \textit{orthonormal} basis. Given a metric tensor $g$, the dreibein basis is given by $\{ e_a = e_a^{\ \mu} e_\mu \}$ with corresponding dual basis $ \{ e^a = e^a_{\ \mu} e^\mu \}$, which satisfies $g(e_a , e_b) = \eta_{ab}$ and $e^a(e_b) = \delta^a_b$, where $\eta_{ab} = \mathrm{diag}(1,-1,-1)$ is the Minkowski metric. In components, these relations read
\begin{equation}
g_{\mu \nu} e_a^{\ \mu} e_b^{\ \nu} = \eta_{ab}, \quad e^a_{\ \mu} e_b^{\ \mu} = \delta^a_b. \label{eq: dreibein def}
\end{equation}
We use \textit{Latin} indices ranging over $0,1,2$ to represent components of tensors with respect to the dreibein basis.

The components of the dreibein $e_a^{\ \mu}$ themselves are sometimes called the dreibein, while the components of the dual dreibein $e^a_{\ \mu}$ are called the \textit{inverse dreibein} as they allow one to invert the expressions (\ref{eq: dreibein def}). However, we adopt the convention of simply calling them both the dreibein. The dreibein also allow us to transform components between frames, i.e. for a $(1,0)$ tensor $A$ we have $A^\mu = e_a^{\ \mu} A^a$ and $A^a = e_{\ \mu}^a A^\mu$ and so on for higher rank tensors. The metrics $g_{\mu \nu}$ and $\eta_{ab}$ and their inverses $g^{\mu \nu}$ and $\eta^{ab}$ allow us to raise and lower Greek and Latin indices, respectively.
\subsubsection{Spin Connection}
The covariant derivative of a tensor is usually expressed in terms of the coordinate basis. For example, the covariant derivative of a rank $(1,1)$ tensor $A^\mu_{\ \nu}$ is given by
\begin{equation}
\nabla_\alpha A^\mu_{\ \nu} = \partial_\alpha A^\mu_{\ \nu} + \Gamma^\mu_{\ \alpha
 \beta } A^\beta_{\ \nu} - \Gamma^\beta_{\  \alpha \nu } A^\mu_{\ \beta}, \label{eq:covariant derivative}
\end{equation}
where $\Gamma^\alpha_{\ \beta \gamma}$ are the components of the \textit{connection}. From this, we can extend the definition to tensors of arbitrary rank with a factor of $\Gamma^\alpha_{\ \beta \gamma}$ for each index of the tensor. 

The choice of our connection defines what our covariant derivative is. It is commonplace to restrict to a \textit{metric compatible} connection, that is, a connection for which $\nabla g = 0$. It can be shown that a metric compatible connection takes the form
\begin{equation}
\Gamma^\rho_{\ \mu \nu} = \tilde{\Gamma}^\rho_{\ \mu \nu} + K^\rho_{\ \mu \nu}, \label{eq: metric connection}
\end{equation}
where $\tilde{\Gamma}^\rho_{\ \mu \nu}$ is the \textit{Levi-Civita} connection, sometimes called the \textit{Christoffel symbols}, and $K^\rho_{\ \mu \nu}$ is the contortion tensor. The Levi-Civita connection is completely determined by the metric
\begin{equation}
\tilde{\Gamma}^\rho_{\ \mu \nu} = \frac{1}{2} g^{\rho \sigma}(\partial_\mu g_{\sigma \nu} + \partial_\nu g_{\sigma \mu} - \partial_\sigma g_{\mu \nu}), 
\label{eq: levi-civita def}
\end{equation}
and is symmetric on exchange of $\mu$ and $\nu$. On the other hand, the contortion tensor is given by
\begin{equation}
K^\rho_{\ \mu \nu} = \frac{1}{2}( T^\rho_{\ \mu \nu} + T_{\mu \ \nu}^{\ \rho} + T_{\nu \ \mu}^{\ \rho}). \label{eq: contortion def}
\end{equation}
where $T^\rho_{\ \mu \nu} = 2 \Gamma^\rho_{[\mu \nu]}$ is the torsion tensor corresponding to the connection $\Gamma^\rho_{\ \mu \nu}$. Square brackets denote anti-symmetrisation over the indices contained within the bracket. 

The tools introduced so far can be expressed in terms of the dreibein basis. The components of the \textit{spin connection} are given by
\begin{equation}
\omega_{\mu b}^a = e^a_{\ \alpha}(\partial_\mu e^{\ \alpha}_b + \Gamma^\alpha_{\ \mu \beta} e^{\ \beta}_b). 
\label{eq: spin connection 2}
\end{equation}
In this case, analogous to (\ref{eq:covariant derivative}), the covariant derivative of a $(1,1)$ tensor $A^a_{\ b}$ in the dreibein basis will be
\begin{equation}
\nabla_\mu A^a_{\ b} = \partial_\mu A^a_{\ b} + \omega_{\mu c}^a A^c_{\ b} - \omega_{\mu b}^c A_{\ c}^a.
\end{equation}
The spin connection is of interest to us as it allows one to take covariant derivatives of spinors which will be discussed in more detail in the next section.

If we work with a metric compatible connection given by (\ref{eq: metric connection}), we see that the spin connection will expand out as
\begin{equation}
\omega_{\mu b}^a = \tilde{\omega}_{\mu b}^a + K_{ \ \mu b}^a, \label{eq: LC spin connection}
\end{equation}
where $\tilde{\omega}_{\mu b}^a$ is the Levi-Civita spin connection related to the coordinate Levi-Civita connection via the formula (\ref{eq: spin connection 2}) with $\Gamma^\alpha_{\  \mu \beta}$ replaced with $\tilde{\Gamma}^\alpha_{\ \mu \beta}$ and $K_{\ \mu b}^a = e^a_{\ \alpha} e^{\ \beta}_b K^\alpha_{\   \mu \beta}$ is the contortion tensor simply expressed in the dreibein basis instead. It can be shown that for a metric compatible connection, if we treat the dreibein $e_a^{\ \mu}$ as the components of a $(1,1)$ tensor, then $\nabla_\mu e^{\ \nu}_a = 0$, which is known as the \textit{dreibein postulate}.


\subsubsection{Curvature and Torsion}
\label{sec:curvature}
The connection of a space also defines two extra geometric quantities, the curvature and the torsion. The torsion of a connection has already been defined as
\begin{equation}
T^\rho_{\ \mu \nu} = 2\Gamma^\rho_{[\mu \nu]}, 
\label{eq: torsion}
\end{equation}
with respect to the coordinate basis. As the Levi-Civita connection $\tilde{\Gamma}^\rho_{\mu \nu}$ is symmetric under interchange of $\mu $ and $\nu$, its corresponding torsion vanishes. The curvature of a connection is given by the Riemann tensor defined as
\begin{equation}
R^\rho_{\  \sigma \mu \nu} = \partial_\mu \Gamma^\rho_{\ \nu \sigma} - \partial_\nu \Gamma^\rho_{\ \mu \sigma} + \Gamma^\rho_{\ \mu \lambda} \Gamma^\lambda_{\ \nu \sigma} - \Gamma^\rho_{\ \nu \lambda}\Gamma^\lambda_{\ \mu \sigma} .
\label{eq: curvature}
\end{equation}
From the Riemann tensor, we can obtain two more geometric quantities, the Ricci tensor defined as $R_{\mu \nu} = R^\sigma_{\ \mu \sigma \nu}$ and the Ricci scalar defined as $R = R^\mu_{\ \mu}$. 

In terms of the full connection $\Gamma^\alpha_{\mu \nu} = \tilde{\Gamma}^\rho_{\ \mu \nu} + K^\rho_{\ \mu \nu}$ the Riemann tensor is given by
\begin{equation}
\begin{aligned}
R^\rho_{\ \sigma \mu \nu}  = &\,\, \tilde{R}^\rho_{\ \sigma \mu \nu} + 2 \partial_{[ \mu} K^\rho_{\ \nu] \sigma}  + 2 \tilde{\Gamma}^\rho_{[\mu | \lambda} K^\lambda_{\ \nu] \sigma} \\
&  + 2K^\rho_{\ [\mu | \lambda} \tilde{\Gamma}^\lambda_{\nu]  \sigma} + 2K^\rho_{\ [ \mu | \lambda } K^\lambda_{\ \nu ] \sigma},
\end{aligned}
\end{equation}
where $ \tilde{R}^\rho_{\ \sigma \mu \nu}$ is defined equivalently to $R^\rho_{\ \sigma \mu \nu}$ in \rf{eq: curvature}, but with $\Gamma^\alpha_{\ \beta \mu}$ replaced with $\tilde{\Gamma}^\alpha_{\ \beta \mu}$. The bracket notation $A_{[\mu| \nu} B_{\rho]}$ for example denotes anti-symmetrisation over only $\mu$ and $\rho$, leaving $\nu$ alone. The corresponding Ricci scalar is given by 
\begin{equation}
R  = \tilde{R} - K_{\rho \mu \nu} K^{\rho \mu \nu},
\label{eqn:totalcurvature}
\end{equation}
where we have assumed that the contortion is completely anti-symmetric. We shall employ this simple formula in order to determine the curvature of the spaces we consider.


\subsection{Spinor Fields on Riemann-Cartan Geometry}

\subsubsection{The Dirac Action}
The action for a spin-$\frac{1}{2}$ particle $\psi$ of mass $m$ defined on a general $(2+1)$-dimensional Riemann-Cartan spacetime $M$ is given by\cite{Nakahara}
\be
S_\text{RC}  = \frac{i}{2} \int_M \mathrm{d}^{2+1} x  |e| \left(  \bar{\psi} \gamma^\mu D_\mu \psi - \overline{D_\mu \psi} \gamma^\mu \psi + 2i m \bar{\psi}\psi \right), \label{eq:dirac action}
\ee
where $\{ \gamma^\mu \}$ are the \textit{curved} space gamma matrices. These matrices obey the Clifford algebra $\{ \gamma^\mu , \gamma^\nu \} = 2 g^{\mu \nu}$ and are related to the flat space gamma matrices $\{ \gamma^a \}$ via $\gamma^\mu = e^{\ \mu}_a \gamma^a$, which obey the flat space Clifford algebra $\{ \gamma^a, \gamma^b \} = 2\eta^{ab}$. The gamma matrices obey $(\gamma^a)^\dagger = \gamma^0 \gamma^a \gamma^0$. The object $|e| = | \det [e^a_{\ \mu}] |$ which from (\ref{eq: dreibein def}) obeys $ |e| = \sqrt{|g|}$, where $g$ is the determinant of the metric. Using the flat space gamma matrices, we define the \textit{Dirac adjoint} $\bar{\psi} = \psi^\dagger \gamma^0$.

The covariant derivative of spinors is given by
\begin{align}
D_\mu \psi & = \partial_\mu \psi + \omega_\mu \psi, \\
\overline{D_\mu \psi} & = (D_\mu \psi)^\dagger \gamma^0 = \partial_\mu \bar{\psi} - \bar{\psi} \omega_\mu,
\end{align}
where $\omega_\mu$ is given by
\begin{equation}
\omega_\mu = \frac{i}{2} \omega_{\mu a b} \Sigma^{ab}, \quad \Sigma^{ab} = \frac{i}{4}[\gamma^a,\gamma^b],
\end{equation}
and $\omega_{\mu a b} = \eta_{ac} \omega_{\mu b}^c$ are the components of the spin connection. The operators $\{ \Sigma^{ab} \}$ are the generators of the Lorentz algebra $\mathfrak{so}(2,1)$. We use the notation $D_\mu$ instead of $\nabla_\mu$ to highlight the fact these derivatives are acting on spinors and not tensors. In this paper we shall refer to $\omega_\mu$ as the connection as well.

When quantising this theory, we impose the curved space fermionic anti-commutation relations
\begin{equation}
\begin{aligned}
\{ \psi_\alpha(t,\boldsymbol{x}), \psi_\beta(t,\boldsymbol{x}') \} & = \{ \psi^\dagger_\alpha(t, \boldsymbol{x}), \psi_\beta^\dagger(t,\boldsymbol{x}')\} = 0, \\ 
\{ \psi^\dagger_\alpha(t,\boldsymbol{x}), \psi_\beta(t,\boldsymbol{x}')\} & = \frac{i}{|e|} \delta_{\alpha \beta} \delta^{(2)}(\boldsymbol{x}-\boldsymbol{x}'), \label{eq:commutator}
\end{aligned}
\end{equation}
where $\alpha,\beta$ label the components of the spinors and $\delta^{(2)}(\boldsymbol{x}-\boldsymbol{x}')$ is the two-dimensional Dirac delta.

It is important to note that the continuum limit of the KHLM can be described using a single-particle Hamiltonian (\ref{eq:DiracKit}) expressed with respect to a spinor field $\Psi$ obeying \textit{flat} spacetime anti-commutation relations. These flat spacetime anti-commutation relations are simply (\ref{eq:commutator}) with $|e|=1$. In order to compare the lattice and quantum field theory Hamiltonians at the single particle level, the corresponding spinors need to satisfy the same anti-commutation relations. Hence, we must renormalise the spinors $\psi$ of the Riemann-Cartan theory in (\ref{eq:dirac action}) in order to obey the flat space anti-commutation relations. This is achieved by defining 
\begin{equation}
\chi = \sqrt{|e|} \psi, \label{eq:spinor}
\end{equation} 
which indeed obeys $\{ \chi^\dagger_\alpha(t,\boldsymbol{x}), \chi_\beta(t,\boldsymbol{x}')\} = {i} \delta_{\alpha \beta} \delta^{(2)}(\boldsymbol{x}-\boldsymbol{x}')$, the flat spacetime anti-commutation relations. We can then make the identification $\Psi = \chi$ between the spinors of the two theories.

With this observation, we substitute our new spinor $\chi$ into the Dirac action (\ref{eq:dirac action}). If we explicitly expand out the covariant derivatives, we have
\begin{equation}
S_\text{RC} = \frac{i}{2} \int_M \mathrm{d}^{2+1}x  \left(  \bar{\chi} \gamma^\mu \partial_\mu \chi - \partial_\mu \bar{\chi} \gamma^\mu \chi + \bar{\chi} \{ \gamma^\mu, \omega_\mu \} \chi \right),
\end{equation}
where to declutter the algebra we have temporarily set $m=0$. We now integrate by parts to remove $\partial_\mu \bar{\chi}$. This gives 
\begin{equation}
S_\text{RC} = \int_M \mathrm{d}^{2+1} x  \bar{\chi}\left( i  \gamma^\mu \partial_\mu + \frac{i}{2} \{ \gamma^\mu, \omega_\mu \} + \frac{i}{2}  \partial_\mu \gamma^\mu \right) \chi.
\end{equation}
Using the identity that in $(2+1)$-dimensional spacetime we have $\{ \gamma^a, [ \gamma^b, \gamma^c ] \} = 4 \epsilon^{abc} \gamma^0 \gamma^1 \gamma^2$, where $\epsilon^{abc}$ is the Levi-Civita symbol, we can simplify the second term in the integrand as
\begin{equation}
\{ \gamma^\mu, \omega_\mu \} = - \frac{1}{8} \omega_{\mu ab} \{ e_c^{\ \mu} \gamma^c , [\gamma^a, \gamma^b]\} = -\frac{1}{2} \omega_{abc} \epsilon^{abc} \gamma^0 \gamma^1 \gamma^2,
\end{equation}
where $\omega_{abc} = e_a^{\ \mu} \omega_{\mu bc}$. The final form of the action is given by
\begin{equation}
S_\text{RC} = \int_M \mathrm{d}^{2+1} x \bar{\chi} \left( i \gamma^\mu \partial_\mu - \frac{i}{4} \omega_{abc}\epsilon^{abc} \gamma^0 \gamma^1 \gamma^2 + \frac{i}{2} \partial_\mu \gamma^\mu \right) \chi. \label{eq:dirac action 2}
\end{equation}


\subsubsection{The Hamiltonian}
We restrict our spacetime $M$ to be a static spacetime of the form 
\be
M = \mathbb{R} \times \Sigma
\ee 
with the natural coordinate system $(t,x^i)$, where $\mathbb{R}$ corresponds to time and $\Sigma$ is a two-dimensional curved space. In this way, we are assuming that only the purely spatial part of spacetime is curved and time can be viewed as a parameter. This is the case that corresponds to the geometric description of KHLM as time remains unaffected by the distortion of the system's couplings. 

The metric tensor for $M$ takes a block-diagonal form in the natural coordinate system
\begin{equation}
g_{\mu \nu} = \begin{pmatrix} 1 & 0 & 0 \\ 0 & g_{xx} & g_{xy} \\ 0 & g_{xy} & g_{yy} \end{pmatrix} 
\label{eq:metric}
\end{equation}
and will be a constant in coordinate time, $\partial_t g_{\mu \nu} = 0$. The dreibein that correspond to this metric will take the form
\begin{equation}
e^a_{\ \mu} =  \begin{pmatrix} 1 & 0 & 0 \\ 0 & e^1_{\ x} & e^1_{\ y} \\ 0 & e^2_{\ x} & e^2_{\ y} \end{pmatrix} , \quad e_a^{\ \mu} = \begin{pmatrix} 1 & 0 & 0 \\ 0 & e_1^{\ x} & e_1^{\ y} \\ 0 & e_2^{\ x} & e_2^{\ y} \end{pmatrix},\label{eq:dreibein}
\end{equation}
where the convention taken is that the index $a$ runs down the columns while the index $\mu$ runs along the rows. 

We make two observations which allow us to simplify the action (\ref{eq:dirac action 2}). First, using the definition (\ref{eq: levi-civita def}), we see that any time components of the Levi-Civita connection $\tilde{\Gamma}^\rho_{\ \mu \nu} $ will vanish: $\tilde{\Gamma}^t_{\ \mu \nu} =\tilde{\Gamma}^\rho_{\ t \nu} = \tilde{\Gamma}^\rho_{\  \nu t}= 0$. This means the time components of the corresponding Levi-Civita spin connection will also vanish: $\tilde{\omega}^a_{t b} = \tilde{\omega}^0_{ \mu b} = \tilde{\omega}^a_{ \mu 0} = 0 $, while the contortion of the spin connection remains unaffected. If we expand out the spin connection term in the action (\ref{eq:dirac action 2}) we have
\begin{equation}
\omega_{abc} \epsilon^{abc} = \tilde{\omega}_{abc}\epsilon^{abc} +K_{abc} \epsilon^{abc} =  \frac{1}{2} T_{abc}\epsilon^{abc}, \label{eq:action spin connection}
\end{equation}
so the Levi-Civita connection falls out, where we have used the fact that $\tilde{\omega}_{abc}\epsilon^{abc}=0$ on a static spacetime and replaced the contortion with the torsion using definition (\ref{eq: contortion def}). 

The second observation we can make is by noting that, due to (\ref{eq:action spin connection}), the spinor field only couples to the completely anti-symmetric part of the torsion $T_{abc}$. For this reason, without loss of generality, we take our torsion to be completely anti-symmetric
\begin{equation}
T_{abc} = \frac{1}{3!} \phi \epsilon_{abc},
\end{equation}
where we refer to $\phi$ as the \textit{torsion pseudoscalar}.

With these two observations, the action reduces to the simple form
\begin{equation}
\begin{aligned}
S_\text{RC} & = \int_M \mathrm{d}^{2+1}x \bar{\chi} \left( i \gamma^\mu \partial_\mu - \frac{i}{8} \phi \gamma^0 \gamma^1 \gamma^2 + \frac{i}{2} \partial_i \gamma^i \right) \chi \\
& \equiv \int_M \mathrm{d}^{2+1}x \mathcal{L}_\text{RC},
\label{eq:action}
\end{aligned}
\end{equation}
where $\mathcal{L}_\text{RC}$ is the Lagrangian density.

As we have assumed our spacetime is static, we can define the Hamiltonian from the Lagrangian density via a Legendre transformation as
\begin{equation}
H_\text{RC} = \int_\Sigma \mathrm{d}^2 x \left( \frac{\partial \mathcal{L}_\text{RC}}{\partial \dot{\chi}} \dot{\chi} - \mathcal{L}_\text{RC} \right) \equiv \int_\Sigma \mathrm{d}^2 x \chi^\dagger h_\text{RC} \chi.
\end{equation}
The single-particle Hamiltonian $h_\text{RC}$ is given by
\begin{equation}
h_\text{RC} = e_a^{\ i} \gamma^0 \gamma^a p_i + \frac{i}{8} \phi \gamma^1 \gamma^2  + \frac{i}{2} \partial_i e_a^{\ i} \gamma^0 \gamma^a + m \gamma^0, 
\label{eq:finalHam}
\end{equation}
where we have written it down explicitly in terms of the dreibein and reinstated the mass $m$. The canonical momentum operator is given by $p_i = -i\partial_i$.

\section{Riemann-Cartan geometry from the Kitaev honeycomb lattice model \label{Riemann-Cartan KHLM}}

We now deform the original Kitaev model by varying its couplings in order to obtain a Riemann-Cartan Hamiltonian in the continuum limit, as given by \rf{eq:finalHam}. There are several aspects of this Hamiltonian that we would like to identify. 

The non-trivial geometry of a Riemann-Cartan theory is encoded in the dreibein $ e_a^{\ \mu} $ and torsion pseudoscalar $\phi$. These objects arise in the Hamiltonian with their own respective terms which both require corresponding terms in the microscopic model to emerge. The same applies to the mass $m$. In addition to this, we require a metric that has space-dependent components in order to achieve a non-trivial curvature. This implies that the parameters $\{ J_i \}$ and $K$ of the micoscropic model have to be upgraded to space-dependent parameters.

In this section, we shall assume that the continuum limit of the model with space-dependent parameters takes the same form as the model with constant parameters, but where the parameters have been simply upgraded to slowly varying space-dependent functions. For this reason, we first take the continuum limit in the constant parameter case in order to extract the dreibein and torsion, and then upgrade them to space-dependent functions to extract the curvature. 

Note that any continuum limit of the KHLM for constant parameters will not yield the $\partial_i e_a^{ \ i }$ term of the general Riemann-Cartan Hamiltonian (\ref{eq:finalHam}), however this is not a problem because all of the important information about the dreibein is contained in the kinetic term.


\subsection{The isotropic $J_x=J_y=J_z=J$ case \label{sec:isotropic}}
We first focus on the isotropic coupling case for which $J_x=J_y=J_z=J$. The corresponding continuum limit for which $J$ is a constant is given by
\begin{equation}
h_\text{KHLM}  = 3J \gamma^0( \gamma^1 p_x -  \gamma^2 p_y ) -  3\sqrt{3}Ki \gamma^1 \gamma^2 .
\label{eq:iso_ham}
\end{equation}
We proceed to interpret this as a Riemann-Cartan Hamiltonian of the form \rf{eq:finalHam} and identify the corresponding dreibein, metric, curvature and torsion of the model.
\subsubsection{The Metric}
A direct comparison of the isotropic continuum limit \rf{eq:iso_ham} with the Riemann-Cartan Hamiltonian \rf{eq:finalHam} reveals that the dreibein of the model are given by
\begin{equation}
e_a^{\ \mu} = \begin{pmatrix}
1 & 0 & 0 \\
0 & 3J & 0 \\
0 & 0 & -3J
\end{pmatrix},
\quad
e^a_{\ \mu} = \begin{pmatrix}
1 & 0 & 0 \\
0 & \frac{1}{3J} & 0 \\
0 & 0 & -\frac{1}{3J}
\end{pmatrix},
\end{equation}
with the corresponding metric
\begin{equation}
g_{\mu \nu} = e^a_{\ \mu}e^b_{\ \nu}\eta_{ab} =
\begin{pmatrix}
1 & 0 & 0 \\
0 & -\frac{1}{9J^2} & 0 \\
0 & 0 & -\frac{1}{9J^2} \label{eq:iso_metric}
\end{pmatrix}.
\end{equation}
We see that the $J$ term alone determines the metric of the model and is unaffected by the $K$ term.

\subsubsection{Curvature and Torsion}
As presented previously, the curvature and torsion of a spacetime depends upon which connection we are working with. We first calculate the curvature of the Levi-Civita connection $\tilde{\Gamma}^\rho_{\ \mu \nu}$.

As we are working on a static spacetime of the form $M = \mathbb{R} \times \Sigma$, the metric, after diagonalisation, takes the form
\begin{equation}
g_{\mu \nu} =
\begin{pmatrix}
1 & 0 & 0 \\
0 & F & 0 \\
0 & 0 & G
\end{pmatrix}, \label{eq:generaldiagmetric}
\end{equation}
where $F = F(x,y)$ and $G=G(x,y)$ are arbitrary functions of space only. A metric of this form yields the Levi-Civita connection $\tilde{\Gamma}^\alpha_{\mu \nu} = \frac{1}{2}g^{\alpha \beta}( \partial_\mu g_{\beta \nu} + \partial_\nu g_{\beta \mu} - \partial_\beta g_{\mu \nu})$, where
\begin{equation}
\begin{aligned}
\tilde{\Gamma}^x_{x x} & = \frac{1}{2F} \partial_x F, \,\,\, \tilde{\Gamma}^x_{x y} = \tilde{\Gamma}^x_{y x} = \frac{1}{2F}\partial_y F, \\
 \tilde{\Gamma}^x_{yy} & = - \frac{1}{2F} \partial_x G, \,\,\, \tilde{\Gamma}^y_{yy}  = \frac{1}{2G} \partial_y G , \\\tilde{\Gamma}^y_{xy} & = \tilde{\Gamma}^y_{yx} = \frac{1}{2G} \partial_x G, \,\,\, \tilde{\Gamma}^y_{xx} = - \frac{1}{2G} \partial_y F,
\end{aligned}
\label{eqn:connections}
\end{equation}
while all of the time components vanish.

As the Levi-Civita connection has no time components and is constant in time, all time components of the corresponding Riemann tensor are zero. For this reason we can restrict ourselves to the two-dimensional spatial hypersurface $\Sigma$. It can be shown that the Riemann tensor in a two-dimensional space has only one independent component and is given by
\begin{equation}
\tilde{R}_{ijkl} = \frac{1}{2}\tilde{R}(g_{ik}  g_{lk} - g_{il}g_{kj}), \label{eq:2D riemann}
\end{equation}
where $\tilde{R}$ is the Ricci scalar and $i,j,k,l$ denote spatial components. With a metric in the form \rf{eq:generaldiagmetric}, giving the connection \rf{eqn:connections},  we find the corresponding Ricci scalar is given by
\begin{equation}
\tilde{R}  = -\frac{1}{2} \bigg[ \partial_x \left({ \partial_x G\over FG}  \right) + \partial_y \left({ \partial_y F\over FG}  \right) + {\partial_x^2 G + \partial _y^2 F  \over FG} \bigg]
\label{eq:Ricci0}
\end{equation}
With the Ricci scalar at hand, the Riemann tensor is fully determined. This is a result we shall refer back to later on in the paper.

For the isotropic case we have $F = G = -\frac{1}{9J^2} $. Employing \rf{eq:Ricci0}, we upgrade the parameter $J$ to an arbitrary function of space which yields the Ricci scalar
\begin{equation}
\tilde{R}  =  2 \partial^2 \ln J,
\end{equation}
where $\partial^2 = g^{\mu \nu} \partial_\mu \partial_\nu $ is the Laplacian operator. Hence, in order to obtain non-zero curvature from the Levi-Civita connection, the coupling constant $J$ of the isotropic KHLM needs to be position dependent. In other words, when $\partial_iJ=0$, then $\tilde R=0$. 

Next, we calculate the curvature and torsion of the total connection $\Gamma^\rho_{\ \mu \nu} = \tilde{\Gamma}^\rho_{\ \mu \nu} + K^\rho_{\ \mu \nu}$. Comparison of \rf{eq:iso_ham} with \rf{eq:finalHam} also reveals that the torsion pseudoscalar $\phi$ and mass $m$ are given by 
\begin{equation}
\phi = -24\sqrt{3}K, \quad m = 0,
\end{equation}
so we see the continuum limit describes \textit{massless} Majorana fermions. From $\phi$, the corresponding components of the torsion and contortion in the dreibein basis are given by
\begin{equation}
T_{abc} = -4\sqrt{3} K \epsilon_{abc}, \quad K_{abc} = -2\sqrt{3} K \epsilon_{abc},
\label{eqn:torsionfield}
\end{equation}
so we see that the next-to-nearest $K$ couplings are responsible for torsion in the continuum limit. The torsion pseudoscalar $\phi$ also determines the \textit{total} Ricci scalar which, using (\ref{eqn:totalcurvature}), is given by
\begin{equation}
R = 2\partial^2 \ln J - 72K^2. \label{eq:totalcurvatureiso}
\end{equation}
Note that the total Ricci scalar $R$ is non-zero even when $\partial_iJ=0$ due to the contribution from $\phi$.

To summarise, the continuum limit of the isotropic KHLM corresponds to freely propagating {\em massless} Majorana fermions on a curved spacetime with torsion. Starting from the lattice model, the nearest-neighbour $J$ terms become the kinetic terms in the continuum limit with non-trivial dreibein, while the next-to-nearest-neighbour $K$ terms become the torsional contribution in the continuum limit. Both of these terms contribute to the total curvature of the model as seen in (\ref{eq:totalcurvatureiso})

Originally, the next-to-nearest $K$ terms were derived from three-spin interactions as given in \rf{eq:Kitaev spin}. These interactions are inserted into the KHLM to generate an energy gap in order to give rise to a well-defined non-Abelian topological phase. Despite the gap, this term is not interpreted as a mass in the continuum limit as it does not generate the mass term of (\ref{eq:finalHam}), but instead it is interpreted as the source of torsion as given in \rf{eqn:torsionfield}.


\subsubsection{Mass from Kekul\'e distortions}
The energy gap of the original Kitaev model comes from the $K$ term that is equivalent to torsion in the continuum limit. To give rise to a mass term in the continuum limit, we need to introduce a Kekul\'e distortion to the tunnelling couplings of the Majorana fermions. This method is similar to the one employed in graphene to theoretically generate a mass gap~\cite{Hou} and adjusted further to the case of Majorana fermions~\cite{Yang}.

\begin{figure}[t]
\center
\includegraphics[width =\linewidth]{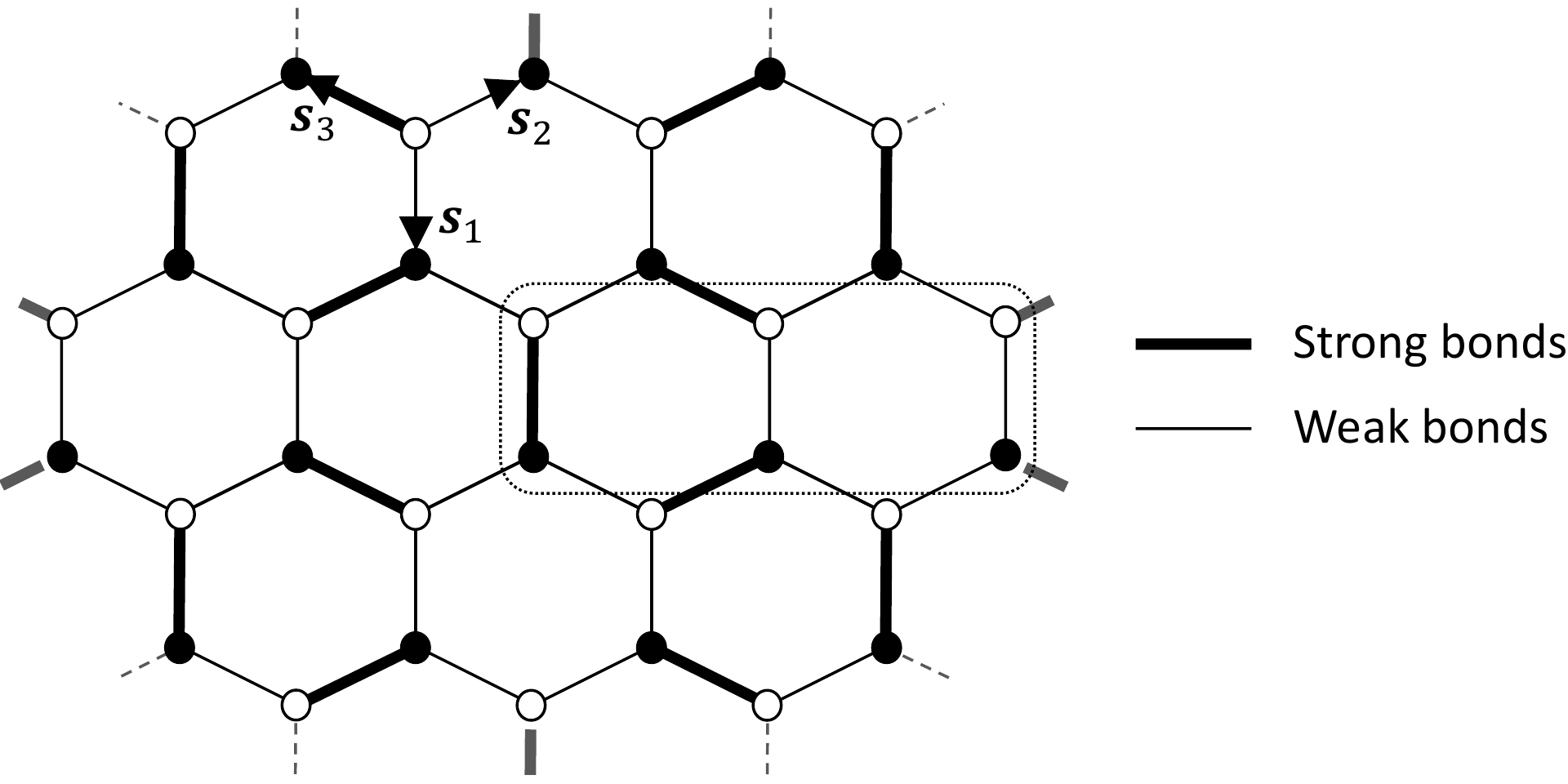}
\caption{The Kekul\'e distortion in the couplings of the honeycomb lattice model, as described by Eqns. \rf{eq:Kekule1} and \rf{eq:Kekule2}, which generate a mass term in the Hamiltonian. Strong and weak tunnelling couplings are indicated as thick and thin bonds, respectively, between lattice sites. This configuration of couplings is periodic with respect to a unit cell with six sites, as shown. The vectors $\boldsymbol{s}_1= (0,-1)$, $\boldsymbol{s}_2 =({\sqrt{3}\over 2},{1\over 2})$, $\boldsymbol{s}_3 = (-{\sqrt{3}\over 2},{1\over 2})$ used in \rf{eq:Kekule2} translate between lattices $A$ and $B$ are also depicted.}
\label{fig:Kekule}
\end{figure}

To proceed we consider an additional term in the lattice Hamiltonian of the Kitaev honeycomb lattice model of the form
\be
\delta H = i \sum_{i\in A} \sum _{k=1}^3 \delta J_{ik} c^a_{{\bs r}_i} c^b_{{\bs r}_i+{\bs s}_k} +\text{h.c.},
\label{eq:Kekule1}
\ee
where $A$ is the sub-lattice defined in Fig.~\ref{fig:honeycomb} and the vectors ${\bs s}_1$, ${\bs s}_2$ and ${\bs s}_3$ are defined in Fig. \ref{fig:Kekule}. With this term the couplings $J_x$, $J_y$ and $J_z$ that were taken to be equal and homogeneous before, are now distorted in the following way:
\be
\delta J_{ik} = {m \over 3} e^{i {\bs P}_+ \cdot {\bs s}_k} e^{i({\bs P}_+-{\bs P}_-)\cdot {\bs r}_i } +\text {c.c.},
\label{eq:Kekule2}
\ee
where $m$ is a constant real number. This additional term causes a Kekul\'e distortion in the couplings of the honeycomb lattice that has the form shown in Fig. \ref{fig:Kekule}. 

The Kekul\'e distortion changes the unit cell of the honeycomb lattice to include six sites, causing the Brillouin zone to fold three times compared to the undisturbed case. Subsequently, we Fourier transform and restrict ourselves to the low energy contributions near the Fermi points.  Up to linear order in momenta the additional term, $\delta H$, gives the following contribution around the Fermi points~\cite{Yang}
\be
\delta H = \Psi^\dagger m \gamma^0 \Psi.
\ee
The contribution to the single-particle Hamiltonian is therefore
\be
h_m =  m\gamma^0,
\ee
where $\gamma^0 = \sigma^x\otimes \mathbb{I}$, which is the mass term given in \rf{eq:finalHam}. Hence, the Majorana fermions in the continuum limit of the KHLM can acquire a mass if a non-trivial Kekul\'e distortion is inserted.

\begin{figure}[t]
\includegraphics[width =\linewidth]{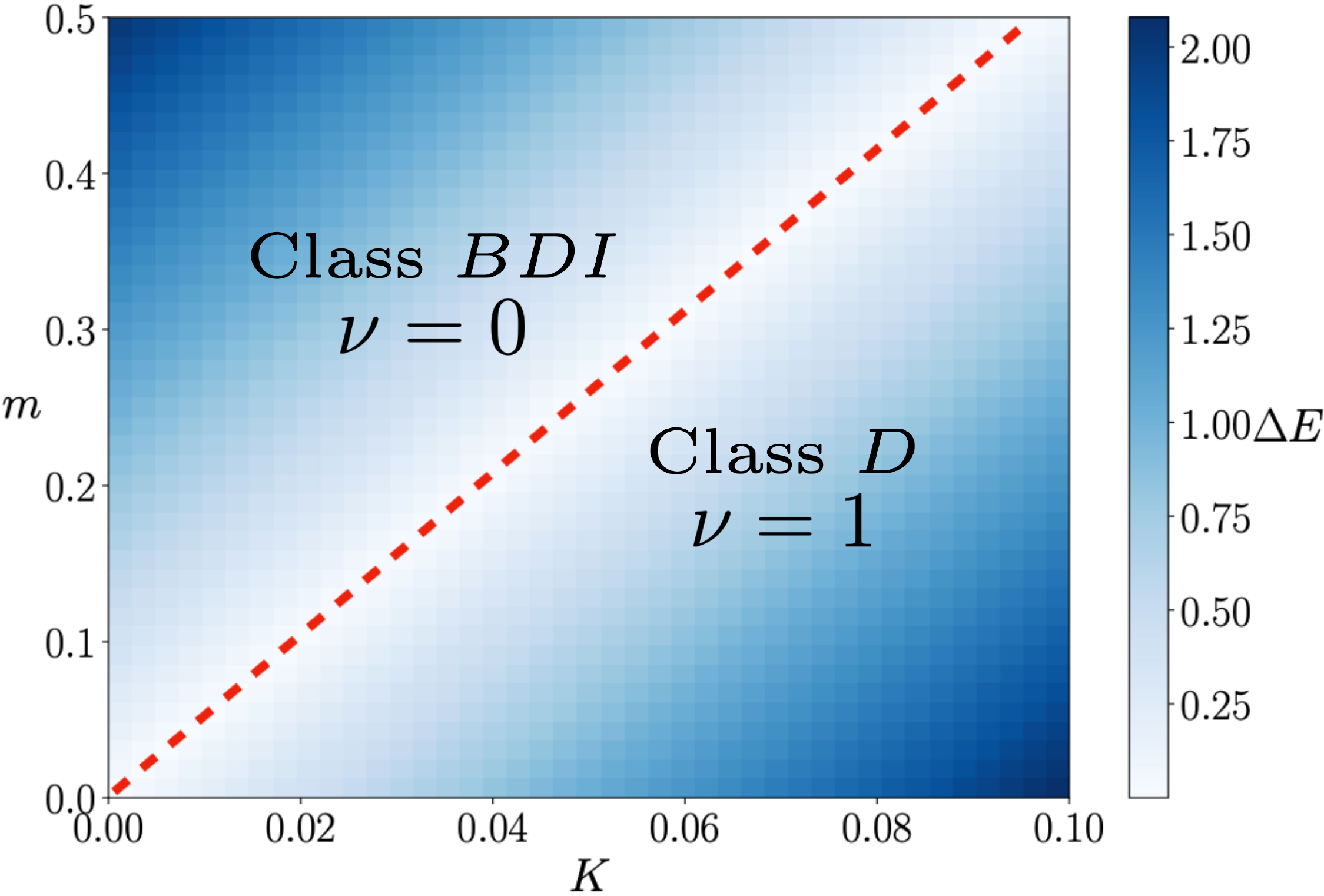}
\vspace*{-0.2cm}
\caption{Phase diagram of the KHLM with its energy gap $\Delta E$ varying as a function of the $K$ coupling and the mass, $m$. By increasing the Kekul\'e distortion a first order phase transition is induced from the gapped topological phase of the KHLM with Chern number $\nu=1$ that belongs in class D to a gapped Kekul\'e phase with Chern number $\nu=0$ that belongs in class BDI. Both of these phases support vortices that bound Majorana zero modes. The red dashed line denotes the analytically obtained phase transition boundary.}
\label{fig:K_delta_trans}
\end{figure}

When $K=0$ the Kekul\'e distortion creates an energy gap due to a non-zero mass $m$. In this situation vertices of sub-lattice $A$ are coupled exclusively to vertices of sub-lattice $B$. So the KHLM recovers its chiral symmetry and the phase of the system belongs in the BDI class that has trivial Chern number~\cite{Chiu}. Nevertheless, zero dimensional defects, such as vortices, can trap chiral Majorana zero modes~\cite{Jackiw,Yang,Chiu}. This should be contrasted with the case where only the $K$-term is present and the model is equivalent to the $p+ip$ superconductor belonging to class D~\cite{Chiu}. 

By varying the couplings of the extended Kitaev model, it is possible to induce a phase transition between the BDI and D phases. The phase diagram of the model as $m$ and $K$ are varied is shown in Fig.~\ref{fig:K_delta_trans}. It is possible to investigate the nature of the phase transition from the quantum field theory description of the model given by the single-particle Hamiltonian (\ref{eq:finalHam}), where $\phi = -24\sqrt{3}K$. As we consider here the homogeneous and isotropic case we have $e^{\ i}_a=\delta^i_a$. Moreover, the phase transition is given when the energy gap is minimum, so we take $p_i=0$ which is exactly at the Fermi points. Observing that $[i \gamma^1\gamma^2, \gamma^0]=0$, we deduce that the phase transition occurs when the coefficients of these two operators are equal, i.e. $m = 3\sqrt{3} K$. This is verified in Fig.~\ref{fig:K_delta_trans} to be in agreement with the numerical modelling. Due to the commutation relation $[i \gamma^1\gamma^2, \gamma^0]=0$ we deduce that the phase transition is first order occurring due to a simple energy level crossing. We have also verified this behaviour of the spectrum numerically.


\subsection{The generally anisotropic $J$ coupling case}
\label{sec:anisotropy}
\begin{figure}[t]
\center
\includegraphics[width =\linewidth ]{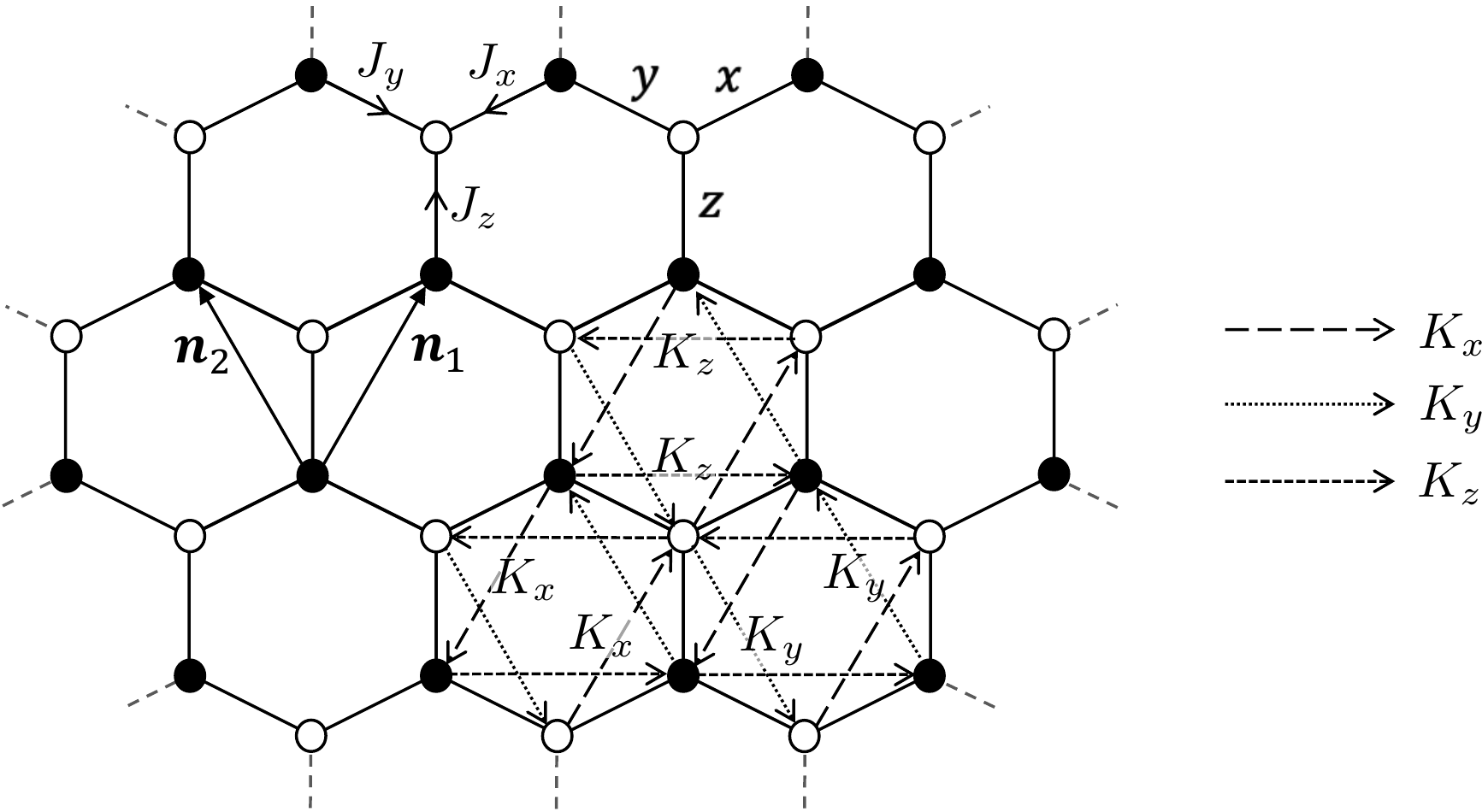}
\caption{The anisotropic KHLM is given by choosing the couplings $J_x$, $J_y$ and $J_z$ to be unequal, giving rise to an anisotropic model. In order to have the $K$-term contribute purely an energy gap we choose the couplings $K_x$, $K_y$ and $K_z$ to be also anisotropic and functions of $J_i$'s, as given by~\rf{eqn:Kcouplings}.
}
\label{fig:honeycombK}
\end{figure}

We now turn away from the isotropic coupling case $J_x=J_y=J_z=J$ and consider the most general metric achievable. In particular, we consider the completely anisotropic case where all $\{J_i\}$ couplings are unequal. Moreover, we allow for anisotropy in the $K$ couplings taking values $K_x$, $K_y$ and $K_z$ depending on their orientation, as shown in Fig.~\ref{fig:honeycombK}. This is to ensure that the Fermi points are not shifted as discussed later in this section. We refer to this case as the generally anisotropic case.

\subsubsection{Continuum Limit}

In this section we take the continuum limit of the generally anisotropic case where the parameters are \textit{constant}.
In this case, the KHLM Hamiltonian takes the form $H = \int \mathrm{d}^2q \psi_\q^\dagger h(\q) \psi_\q $, where $\psi_{\q} = (c^a_{\q} \ ic^b_{\q})^\mathrm{T}$, and the single-particle Hamiltonian $h(\boldsymbol{q})$ given by
\be
h(\q) = 
\begin{pmatrix}
\Delta(\q) &-f(\q) \\
-f^*(\q) & - \Delta(\q) 
\end{pmatrix},
\label{eq:generalham}
\ee
where
\begin{equation}
f({\bs q}) = 2 (J^xe^{i{\bs q} \cdot {\bs n}_1} + J^ye^{i {\bs q} \cdot {\bs n}_2} + J^z),
\label{eq:f_gen}
\end{equation}
and
\begin{equation}
\begin{aligned}
\Delta( \boldsymbol{q})  = 2 \big[ &- K_x \sin( \boldsymbol{q} \cdot \boldsymbol{n}_1) + K_y \sin( \boldsymbol{q} \cdot \boldsymbol{n}_2)  \\
& + K_z \sin( \boldsymbol{q} \cdot ( \boldsymbol{n}_1 - \boldsymbol{n}_2 )) \big].
\end{aligned}
\end{equation}

For $K_i=0$ we expect the system to have Fermi points. Following the procedure we employed in the isotropic case, we solve for the momenta ${\bs q} = {\bs P}$ such that $f({\bs P})=0$. This condition gives two equations
\begin{align}
J_x\cos({\bs P} \cdot {\bs n}_1) + J_y\cos({\bs P}\cdot {\bs n}_2) + J_z & = 0, \label{eq:f_gen_re} \\
J_x\sin({\bs P} \cdot {\bs n}_1) + J_y\sin({\bs P}\cdot {\bs n}_2) & = 0. \label{eq:f_gen_im}
\end{align}
There are two solutions to these equations corresponding to two Fermi points located at momenta
\begin{equation}
\begin{aligned}
{\bs P}_\pm = 
\pm\begin{pmatrix}
\frac{1}{\sqrt{3}}\big(\arccos(a) + \arccos(b)\big) \\
\frac{1}{3} \big(\arccos(a) - \arccos(b)\big)
\end{pmatrix},
\label{eq:gen_fp}
\end{aligned}
\end{equation}
where 
\be
a=\frac{J_y^2-J_x^2-J_z^2}{2J_xJ_z} \,\,\text{and}\,\,
b=\frac{J_x^2-J_y^2-J_z^2}{2J_yJ_z}.
\label{eqn:ab}
\ee
To determine the behaviour of the Hamiltonian around these points, we Taylor expand to first order as
\bq
\!\!\!\!\!\!\!\!\!\!\!f({\bs P}_\pm + {\bs p} ) &=& f({\bs P}_\pm) + {\bs p} \cdot {\bs \nabla} f({\bs P}_\pm) + O(p^2)\nonumber, \\
\!\!\!\!\!\!\!\!\!\!\!\Delta({\bs P}_\pm + {\bs p} ) &=& \Delta({\bs P}_\pm) + {\bs p} \cdot {\bs \nabla} \Delta({\bs P}_\pm) + O(p^2).
\eq
By direct calculation we find
\begin{align}
\begin{split}
\boldsymbol{\nabla} f(\boldsymbol{P}_\pm) &  = 2i \Big[  J_x \left(a \pm i \sqrt{1 - a^2}\right) \boldsymbol{n}_1  \\
&  +J_y\left(b \mp i \sqrt{1 - b^2} \right) \boldsymbol{n}_2 \Big], 
\end{split} \\
\begin{split}
\boldsymbol{\nabla}\Delta (\boldsymbol{P}_\pm )  & = 2 \Big[ -K_x a \boldsymbol{n}_1 + K_y b \boldsymbol{n}_2 \\
 + K_z & \left(ab - \sqrt{1-a^2} \sqrt{1 - b^2}\right) (\boldsymbol{n}_1 - \boldsymbol{n}_2)\Big]. \label{eq:DeltaGeneral1}
\end{split}
\end{align}
For the isotropic case addressed earlier in section \ref{sec:isotropic}, we note that in the continuum limit $f( \boldsymbol{p})$ became the kinetic term whist $\Delta(\boldsymbol{p})$ simply provided an energy gap at the Fermi points. If we demand that this is the case for generally anisotropic couplings, we must ensure that $\Delta(\boldsymbol{p})$ does not shift the position of the Fermi points determined by $f(\boldsymbol{p})$. This can be achieved if \rf{eq:DeltaGeneral1} vanishes. In order to impose this, we must constrain the couplings $\{ K_i \}$ to take the values
\begin{equation}
\begin{aligned}
K_x & = 4K b \left(ab - \sqrt{1-a^2}\sqrt{1-b^2}\right),\\
K_y & =  4K a \left(ab - \sqrt{1-a^2}\sqrt{1-b^2}\right), \\
K_z & = 4K ab,
\label{eqn:Kcouplings}
\end{aligned}
\end{equation} 
where $K \in \mathbb{R}$ is a constant that gives the overall scale of the $K$ couplings and $a$, $b$ are given in \rf{eqn:ab}. The factor of $4$ ensures the gap for the most general case agrees with the gap from the isotropic case in Section \ref{KHLM}. 

With these conditions, the energy gap at each Fermi point is given by $2|\Delta(\boldsymbol{P}_\pm)|$, where $\Delta(\boldsymbol{P}_\pm) = \pm \Delta$ and
\begin{equation}
\Delta =  8 K \sqrt{(1-a^2)(1-b^2)} \big( a \sqrt{1-b^2} + b \sqrt{1-a^2} \big).  
\label{eq:DeltaGeneral}
\end{equation}
At the isotropic limit, $J_x=J_y=J_z=J$, we have from~\rf{eqn:ab} that $a=b=-1/2$ and we recover the corresponding $\Delta = -3\sqrt{3} K$, in agreement with~\rf{eq:4by4H}.

As usual, we expand our Hamiltonian about the Fermi points and define the continuum limit Hamiltonian as $h_\pm ( \boldsymbol{p}) = h(\boldsymbol{P}_\pm + \boldsymbol{p}) $.
We consider the Fermi points simultaneously by defining the four component spinor $\Psi = (c^a_+ \ ic^b_+ \ i c^b_- \ c^a_-)^\mathrm{T}$ where $c^{a/b}_\pm(\boldsymbol{p}) = c^{a/b}_{\boldsymbol{P}_\pm + \boldsymbol{p}}$. We combine the Hamiltonians $h_+(\boldsymbol{p})$ and $h_-(\boldsymbol{p})$ by taking their direct sum with respect to the basis defined by $\Psi$. This yields the total $4 \times 4$ Hamiltonian
\begin{equation}
\begin{aligned}
h_\text{KHLM}  = &\left( A \sigma^z \otimes \sigma^x + B \sigma^z \otimes \sigma^y \right)p_x \\
& + C \sigma^z \otimes \sigma^y p_y + \Delta \mathbb{I} \otimes \sigma^z,
\end{aligned}
\end{equation}
where we have defined the quantities
\begin{subequations}\label{eq:gen_ham_constants}
\begin{align}
A & = \sqrt{ 12J_x^2 - 3 \frac{(J_y^2-J_x^2 - J_z^2)^2}{J_z^2}}, \\
B & =  \sqrt{3} \frac{(J_y^2 - J_x^2)}{J_z^2}, \\
C & = - 3J_z. 
\end{align}
\end{subequations}
This reduces to the original isotropic continuum limit given in~\rf{eq:4by4H} when $J_x = J_y = J_z = J$. In terms of the gamma matrices defined in (\ref{eq:gamma_matrices}), we see that the continuum limit Hamiltonian is given by
\begin{equation}
h_\text{KHLM} = \left(A \gamma^0 \gamma^1 + B \gamma^0 \gamma^2 \right) p_x + C\gamma^0 \gamma^2 p_y + i \Delta  \gamma^1 \gamma^2, \label{eq:gen_aniso_ham}
\end{equation}
which is the most general continuum limit Hamiltonian in Riemann-Cartan form.
\subsubsection{The Metric}
From previously, the general Riemann-Cartan Hamiltonian is given by
\begin{equation}
h_\text{RC} = e_a^{\ i} \gamma^0 \gamma^a p_i + \frac{i}{8} \phi \gamma^1 \gamma^2  + \frac{i}{2} \partial_i e_a^{\ i} \gamma^0 \gamma^a + m \gamma^0 \label{eq:finalHam2}
\end{equation}
Direct comparison of the continuum limit (\ref{eq:gen_aniso_ham}) with the Riemann-Cartan Hamiltonian (\ref{eq:finalHam2}) for constant parameters reveals the dreibein of the model are given by
\begin{equation}
e_a^{\ \mu} = 
\begin{pmatrix}
1 & 0 & 0 \\
0 & A & 0 \\
0 & B & C 
\end{pmatrix},
\quad e^a_{\ \mu} =
\begin{pmatrix}
1 & 0 & 0 \\
0 & \frac{1}{A} & 0 \\
0 & -\frac{B}{AC} & \frac{1}{C}
\end{pmatrix},
\end{equation}
with corresponding metric
\begin{equation}
g_{\mu \nu} = e^a_{\ \mu} e^b_{\ \nu} \eta_{ab} =
\begin{pmatrix}
1 & 0 & 0 \\
0 & -\frac{1}{A^2} - \frac{B^2}{A^2C^2} & \frac{B}{AC^2} \\
0 & \frac{B}{AC^2} & -\frac{1}{C^2}
\end{pmatrix}. \label{eq:generalmetric}
\end{equation}
We also identify the torsion pseudoscalar and mass as 
\begin{equation}
\phi = 8 \Delta, \quad m =0. \label{eqn:torsionfinal}
\end{equation}

\begin{figure}[t]
\center
\includegraphics[width =\linewidth]{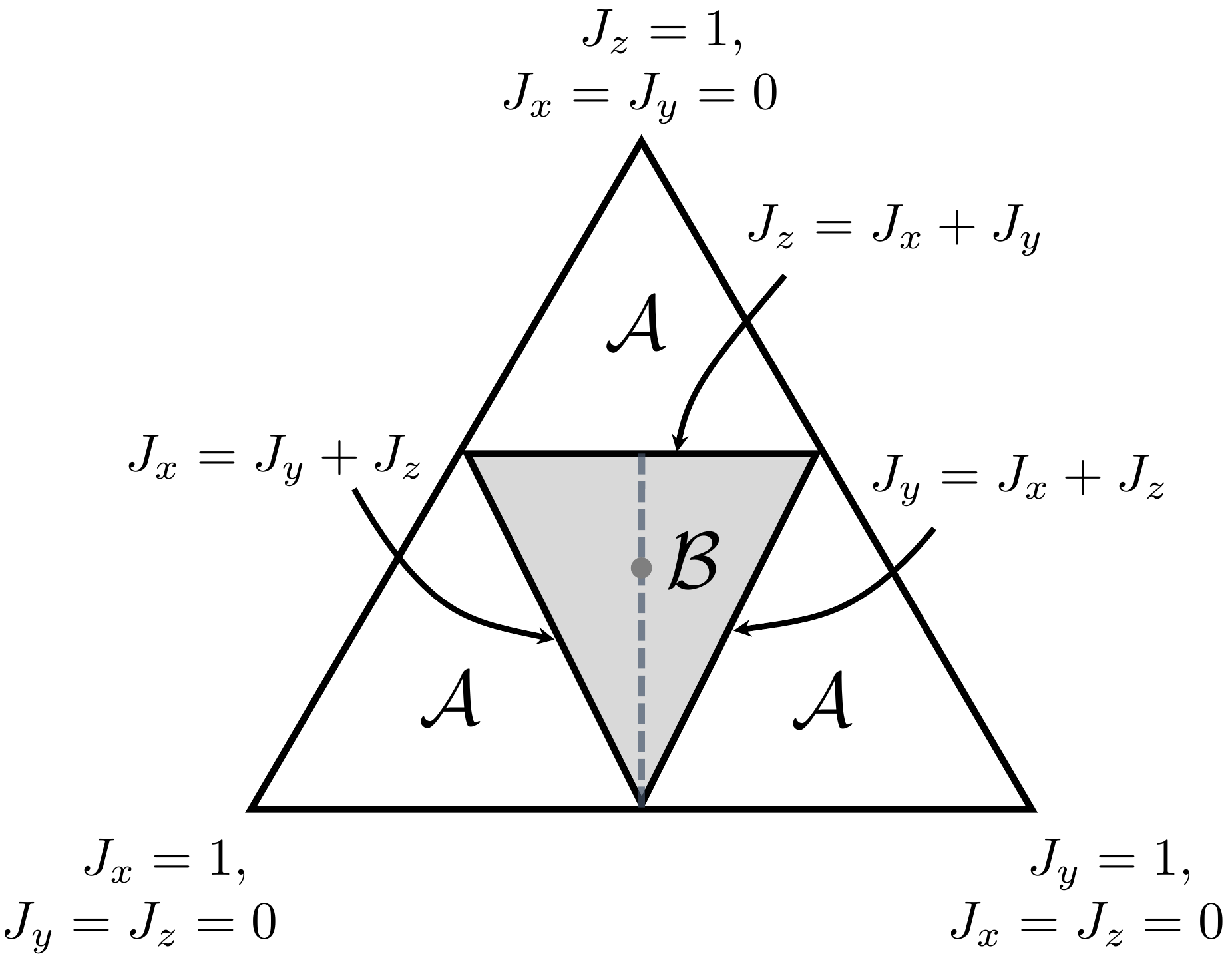}
\caption{The phase diagram of the anisotropic KHLM, where the couplings are normalised as $J_x+J_y+J_z=1$. The isotropic case with $J_x=J_y=J_z$ is denoted by a dot in the centre of the triangles. The quantum spin liquid phase that supports Majorana fermions, denoted as ${\cal B}$, sits in the centre of the diagram. The topological phases ${\cal A}$ correspond to the Toric Code phase. The singularity condition of the metric~\rf{eq:singular} defines the boundaries between the phases ${\cal A}$ and ${\cal B}$. The dashed line corresponds to a possible anisotropic change of couplings in the ${\cal B}$ phase.}
\label{fig:Couplings}
\end{figure}

From (\ref{eq:generalmetric}) we see that the geometry becomes singular when $A=0$ or $C=0$. To analyse the properties of the model when $A=0$ we note that this condition is equivalent to satisfying one of the following equations 
\be
\begin{matrix}
J_x+J_y+J_z = 0, \\
J_x-J_y-J_z = 0, \\
J_x-J_y+J_z = 0, \\
J_x+J_y-J_z = 0.
\end{matrix}
\label{eq:singular}
\ee
Assuming that $J_x$, $J_y$, $J_z$ are all positive we obtain the following conditions that $J_x=J_y+J_z$ or $J_y=J_x+J_z$ or $J_z=J_x+J_y$. Choosing for convenience the normalisation condition $J_x+J_y+J_z=1$ we obtain the triangular phase diagram shown in Fig.~\ref{fig:Couplings}. The above conditions between the $\{ J_i \}$ couplings define the phase boundaries that lie between the non-Abelian phase ${\cal B}$ and the Toric Code\cite{Kitaev2} phases ${\cal A} $ of the model. On the other hand, the condition $C=0$ corresponds to the case where $J_z=0$. This coupling configuration makes the model gapless as it becomes a set of disentangled one-dimensional chains with zero energy gap. The $J_z=0$ case corresponds to the middle of the bottom site of the large triangle in Fig.~\ref{fig:Couplings}. Hence, the geometric description of the KHLM is non-singular within the whole region ${\cal B}$. The singular regions of the metric correspond to the well known phase transitions of the KHLM~\cite{Kitaev}.


\subsection{The anisotropic case with $J_x = J_y = 1$ and $0 \leq J_z \leq 2$}

The phase diagram obtained above corresponds to parameters $\{ J_i \}$ that are constant. We now upgrade these parameters to functions of space to investigate whether the continuum limit can describe a \textit{curved} geometry. For simplicity, we focus our attention on the special anisotropic case where $J_x = J_y = 1$ and $J_z$ can take values between the critical points $0\leq J_z \leq 2$, as shown in Fig.~\ref{fig:Couplings}. We refer to this case as the anisotropic case.
\subsubsection{The Metric}
In the anisotropic case, we see from (\ref{eq:gen_ham_constants}) and (\ref{eq:gen_aniso_ham}) in the previous section that the continuum limit Hamiltonian is given by
\begin{equation}
\begin{aligned}
h_\text{KHLM}  =& \sqrt{12-3J_z^2}\gamma^0 \gamma^1 p_x - 3J_z \gamma^0 \gamma^2 p_y \\
& - KJ_z (4-J_z^2)^\frac{3}{2}i \gamma^1 \gamma^2
\end{aligned}
\end{equation}
Using formula (\ref{eq:generalmetric}), the corresponding metric is given by
\begin{equation}
g_{\mu \nu} = 
\begin{pmatrix}
1 & 0 & 0 \\
0 & \frac{1}{3J_z^2 -12} & 0 \\
0 & 0 & -\frac{1}{9J_z^2} 
\end{pmatrix}.
\label{eqn:metrixexample}
\end{equation}
We see that this agrees with the metric of the isotropic case (\ref{eq:iso_metric}) when $J_z = 1$, in which case this would be describing the isotropic case for $J = 1$.
\subsubsection{Curvature and Torsion}
We now upgrade our coupling constants to slowly-varying functions of space only in order to calculate the curvature. The metric (\ref{eqn:metrixexample}) is diagonal so we can employ the general formula (\ref{eq:Ricci0}) for the Ricci scalar $\tilde{R}$ of the Levi-Civita connection. A direct substitution of 
\begin{equation}
F = \frac{1}{3J_z^2 - 12}, \quad G=-\frac{1}{9J_z^2}
\end{equation}
into the formula yields a rather unpleasant expression for the curvature, but the upshot is that the curvature is non-zero and space-dependent when the coupling constant $J_z$ is space-dependent. The Ricci scalar $\tilde{R}$, together with the formula (\ref{eq:2D riemann}), fully determines the Riemann tensor of the Levi-Civita connection.

Next we determine the torsion of the system. Using equation (\ref{eq:DeltaGeneral}) we find that the gap at the Fermi points is given by
\begin{equation}
\Delta = - K J_z (4-J_z^2)^\frac{3}{2}.
\end{equation}
Combining this result with the equation \rf{eqn:torsionfinal}, we have the torsion pseudoscalar 
\begin{equation}
\phi = - 8KJ_z (4-J_z^2)^\frac{3}{2} \label{eq:torsion_anisotropic}
\end{equation}
which fully determines our torsion $T_{abc} = \frac{1}{3!} \phi \epsilon_{abc}$ and contortion $K_{abc} = \frac{1}{12} \phi \epsilon_{abc}$ tensors. This result, combined with the Levi-Civita connection, fully determines the spacetime connection $\Gamma^\rho_{\ \mu \nu} = \tilde{\Gamma}^\rho_{\ \mu \nu} + K^\rho_{\ \mu \nu}$ of the model. We can use the above results to determine the total Ricci scalar of the model using formula (\ref{eqn:totalcurvature}) which gives
\begin{equation}
R = \tilde{R} - \frac{8}{3}K^2 J_z^2 (4-J_z^2)^3.
\end{equation}

We stated previously that the torsion of the continuum limit is due to the next-to-nearest-neighbour interaction term. Indeed, in the isotropic case the torsion was solely determined by $K$ and had no dependence on the $J$ term---it existed whether we had nearest-neighbour interactions or not. However, despite this fact, we see that the torsion in (\ref{eq:torsion_anisotropic}) actually depends on $J_z$ in this special case. This is due to the fact that we modified our $K$ term to ensure that it did not shift the Fermi points determined by the $J$ term. For this reason, the value of these next-to-nearest neighbour $K$ terms has a dependence on the couplings $\{ J_i \}$, however the overall \textit{scale} of the torsion is determined by $K$ and will vanish if $K=0$. Therefore the identification of the next-to-nearest-neighbour interactions as a source of torsion in the continuum limit remains, albeit the strength being dependent on $J_z$ too.

The same can be said for the total curvature $R$. We see that in the isotropic case, the $J$ term only contributed to the curvature of the Levi-Civita connection and was independent of the contortion. Now, due to the previous reasons, the $J$ term contributes to the total curvature in two ways.

In the next section we shall investigate how faithfully KHLM can reproduce the metric~\rf{eqn:metrixexample} just by varying its couplings.


\section{Geometry of quantum correlations and zero modes}
\label{numerics}

In this section we perform a numerical study of the KHLM with periodic boundary conditions and anisotropic $J$ couplings. Our aim is to determine how faithfully the exact form of the metric obtained in~\rf{eqn:metrixexample} can be reproduced from the lattice description of the microscopic model. To proceed we employ the following procedure. The geometric description of KHLM determines the form of the metric for any configuration of the couplings. By knowing, for example, an eigenstate of the Hamiltonian for a given configuration such as the isotropic case, we can deduce the form of the eigenstate for any other coupling configuration by only considering its spatial transformation according to the corresponding change of the metric. 

To be more concrete consider the transformation of the model from the isotropic case, $J_x=J_y=J_z=1$, to the anisotropic case with $J_x=J_y=1$ and $0\leq J_z\leq 2$ described by the metric~\rf{eqn:metrixexample}.
As $J_z$ varies this metric describes the simultaneous change in the measure of distance for both the $x$ and $y$ directions. To study systematically this effect we consider the case where $J_z$ is constant in space.
In this case the spatial distance $d$ between two points on the distorted space $\Sigma$ is given by
\begin{equation}
d = \sqrt{ - g_{ij} \Delta X^i \Delta X^j },
\end{equation}
where $\Delta X^i$ is the change in coordinates between each point and $g_{ij}$ are the spatial components of the metric. We can visualise what the effect of changing the coupling $J_z$ is on the spatial anisotropy by noting that a unit circle at $J_z=1$ gets deformed into an ellipse as $J_z\neq 1$ with principle axes $d_x$ and $d_y$ along the $x$ and $y$ directions, respectively, satisfying
\begin{equation}
\frac{d_x}{d_y} = \frac{\sqrt{-g_{xx}}}{\sqrt{-g_{yy}}} = \frac{\sqrt{3}J_z}{\sqrt{4-J_z^2}}.
\label{eq:spaceratio}
\end{equation}
In the following we study the effect varying $J_z$ has on physical observables of the model. Specifically we will investigate how faithfully the spatial profiles of two-point Majorana correlations and Majorana zero modes \cite{Kitaev}, bounded at a vortices, are deformed compared to the ratio \rf{eq:spaceratio}.


\begin{figure}[t]
\center
\includegraphics[width=\columnwidth]{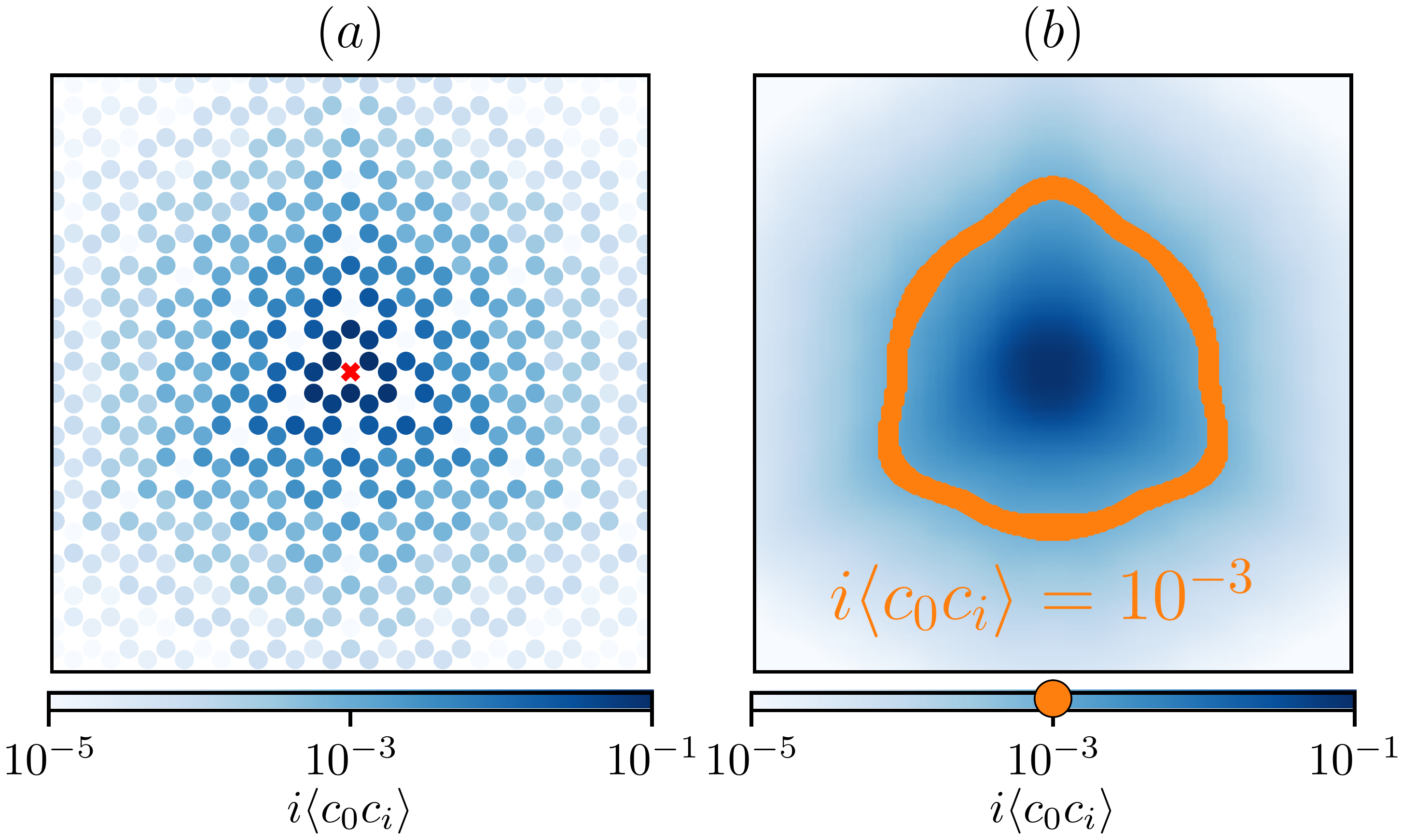}
\caption{ The two-point correlations and their continuous profile. (a) The two-point correlations $i\langle c_0c_i\rangle$ between each site, $i$ and a central reference site, $0$, denoted with a red cross. (b) A continuous approximation of the two-point correlations is constructed using two-dimensional Gaussians centred on each lattice site, as described by~\rf{eqn:shape}. The size and shape of the correlations are characterised by finding the set of points where $i\langle c_0c_i\rangle = 10^{-3}$, as illustrated. We notice that even for large system sizes the hexagonal geometry of the lattice influences the spatial distribution of the correlations. Here we used $J_x=J_y=J_z=1$, system size $36\times36$ $K=0.1$ and $\epsilon = 1$. 
}
\label{fig:lattice-to-bounding-box-corr}
\end{figure}

\subsection{Two-Point Quantum Correlations}

\begin{figure}[t]
\center
\includegraphics[width=\columnwidth]{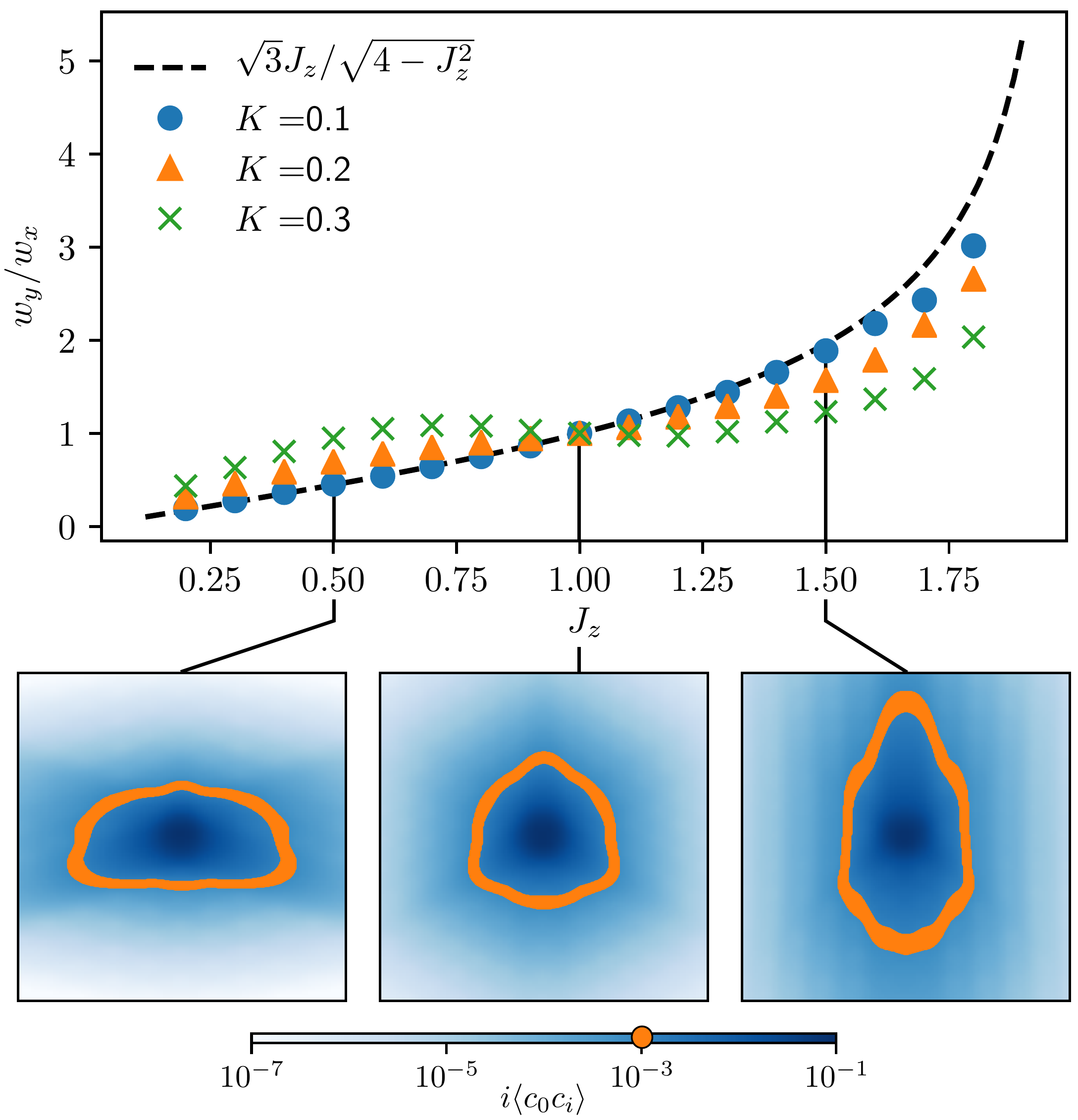}
\caption{Verifying the metric from the continuous approximation of the correlations. The points in the main panel plot the ratio between the height and width of the `boundary' $w_y/w_x$ for $J_x=J_y=1$, $\epsilon =1$, system size $36 \times 36$ and a range of $K$. Also plotted with a dashed line is the theoretically predicted ratio $d_x/d_y = \sqrt{3}J_z/\sqrt{4-J_z^2}$ from Eq.~(\ref{eq:spaceratio}). The numerical data converges to the theoretical line as $K$ decreases. Below are illustrative examples of the boundaries we find at various $J_z$ and $K=0.1$. At the isotropic point, $J_z=1$, we find $w_y/w_x=1$. As $J_z$ deviates from the isotropic point the ratio $w_y/w_x$ can become larger or smaller than one.}
\label{fig:combined-vortex-data-fig-corr}
\end{figure}

Two-point Majorana correlations are the expectation value of a product of two Majorana operators at different sites with respect to the ground state $\ket{\psi_0}$, i.e. $i\langle c_ic_j\rangle = \bra{\psi_0} i c_ic_j\ket{\psi_0}$. They are an important quantity as any other property of the model can be deduced from them as our model is effectively free~\cite{Meichanetzidis}. As the system is gapped we expect the two-point correlations $i\langle c_ic_j\rangle$ to decrease exponentially fast with respect to the distance $|\boldsymbol{r}_i-\boldsymbol{r}_j|$. We extract the two-point correlations by exact diagonalisation of Hamiltonian (\ref{eq:Kitaev}). Taking a single row or column of the correlation matrix gives us a discrete spatial profile of the two-point correlations of each site with respect to a central reference site, as shown in Fig.~\ref{fig:lattice-to-bounding-box-corr}(a).

To study the effect of varying $J_z$, as in Fig.~\ref{fig:Couplings}, on this discrete spatial profile we produce a continuous approximation by replacing the two-point correlation data at each lattice point with a two-dimensional Gaussian centred at the site,
\begin{equation}
i\langle c(\boldsymbol{r})c_j\rangle = \sum_i i\langle c_ic_j\rangle \, \delta(\boldsymbol{r}-\boldsymbol{r}_i) \rightarrow \sum_i \frac{i\langle c_ic_j\rangle}{2\pi\epsilon} e^{-{|\boldsymbol{r}-\boldsymbol{r}_i|^2 \over 2\epsilon}} \, ,
\label{eqn:shape}
\end{equation}
where $\epsilon$ is taken to be of similar magnitude as to the lattice spacing so that the Gaussians of neighbouring sites overlap. Fig.~\ref{fig:lattice-to-bounding-box-corr} illustrates this substitution. This continuous approximation reduces the discrete lattice effects and allows us extract the stretching and squeezing of the observables predicted by (\ref{eq:spaceratio}). It is worth noticing that as the two-point correlations are local they are strongly influenced by the lattice structure of the system. Hence, even if we expected the isotropic point to be rotationally invariant we observe that the honeycomb lattice structure is still evident in the continuum representation even for large system sizes. Nevertheless, this deformation does not obstruct us from demonstrating the equivalence between the microscopic model and the RC geometric theory.

From the continuous approximation of the correlations we numerically identify the set of points where $\langle c(\boldsymbol{r})c_j\rangle$ has decayed to a fixed value e.g. $10^{-3}$. This `boundary' line is drawn for the correlations at the isotropic point of the model in Fig.~\ref{fig:lattice-to-bounding-box-corr}(b). We find that at the isotropic point the boundary is almost circular. As we move away from the isotropic point the boundary should be stretched in either the $x$ or $y$ direction.

To verify the metric (\ref{eqn:metrixexample}) we compare the ratio between the height and width of the boundary $w_y/w_x$ to the ratio of the principle axes of the ellipse $d_x/d_y$, given in (\ref{eq:spaceratio}). We find that at the isotropic point the width and height of the boundary are almost circular, with $w_y/w_x \approx 1$. Fig.~\ref{fig:combined-vortex-data-fig-corr} plots the comparison of $w_y/w_x$ to $d_x/d_y$ for different values of $J_z$. The ratio $w_y/w_x$ converges to $d_x/d_y$ with decreasing $K$. As $K$ decreases the energy gap also decreases, which leads to an increase in the correlation length. When the correlation length becomes large compared to the lattice spacing discrete lattice effects become negligible and the correlations approximate the behaviour of those in a continuous system. We find that these ratios agree very well, particularly away from the phase transition boundaries. We observe that an increase (decrease) in $J_z$ corresponds to a decrease (increase) in the effective distance between lattice sites $d_y$ ($d_x$), resulting in stronger (weaker) correlations along that axis. Thus the correlations appear stretched (squeezed) along that axis of the lattice, i.e. $w_y/w_x$ increases (decreases).


\subsection{Vortex zero-modes}

\begin{figure}[tp]
\center
\includegraphics[width=\columnwidth]{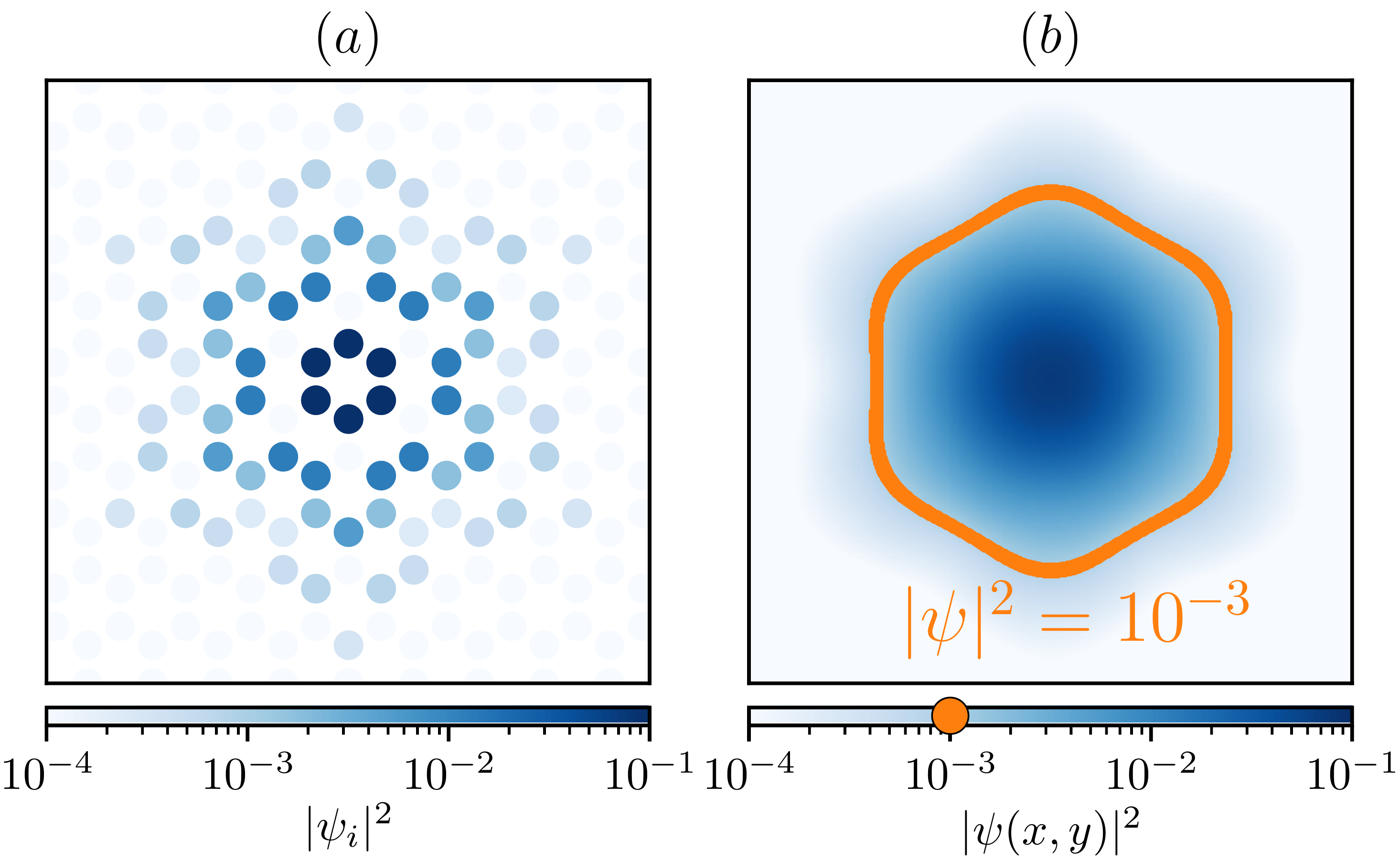}
\caption{Obtaining a continuous profile for the vortex wave function and extracting its dimensions. (a) The discrete lattice probability density $|\psi_i|^2$ of the wave function for a vortex, located on the plaquette in the centre. (b) A continuous approximation of the discrete vortex probability density is constructed using two-dimensional Gaussians centred on each lattice site, as described in the text. The size and shape of the vortex are characterised by finding the set of points where $|\psi(\boldsymbol{r})|^2 = 10^{-3}$, as illustrated. Here we used $J_x=J_y=J_z=1$, system size $36\times36$, $K=0.125$ and $\epsilon = 1$.}
\label{fig:lattice-to-bounding-box}
\end{figure}

Vortex excitations can be introduced in pairs by inserting $\pi$-fluxes into the $\mathbb{Z}_2$ gauge field that couples to the Majorana matter fermions of the model~\cite{Kitaev}. Practically this can be done by flipping the sign of a coupling $J_{ij}$ of a link $(i,j)$ relative to its value in the no-vortex sector. This link belongs to two hexagonal plaquettes, and following this change of gauge, these two plaquettes will each hold a vortex with a Majorana zero-mode localised at each. By subsequent changes in the sign of $J_{ij}$ couplings these two vortices can be moved far apart. In the spectrum of the model the vortex pair manifests as a zero-energy fermion mode~\cite{Kitaev}. We study the spatial wave function of a single vortex sufficiently separated from its pair.
We call the probability density at each lattice site, $|\psi_i|^2$ the discrete spatial profile of the wave function, as shown in Fig.~\ref{fig:lattice-to-bounding-box}(a).

To analyse the geometric profile of the zero modes we adopt the same procedure we used for the Majorana correlations. We approximate the discrete wavefunction profile with a continuous distribution by replacing the probability density at each lattice point with a two-dimensional Gaussian centred at the site,
\begin{equation}
|\psi(\boldsymbol{r})|^2 = \sum_i |\psi_i|^2 \, \delta(\boldsymbol{r}-\boldsymbol{r}_i) \rightarrow \sum_i \frac{|\psi_i|^2}{2\pi\epsilon} e^{-{|\boldsymbol{r}-\boldsymbol{r}_i|^2\over 2\epsilon}} \, ,
\end{equation}
where $\epsilon$ is taken to be similar to the lattice spacing so that the Gaussians of neighbouring sites overlap. Fig.~\ref{fig:lattice-to-bounding-box} illustrates this substitution. In the continuum we expect a wave function exponentially localised at the position of the vortex. This continuous profile reduces the discrete lattice effects allowing us to clearly observe the effects of geometry.

\begin{figure}[tp]
\center
\includegraphics[width=\columnwidth]{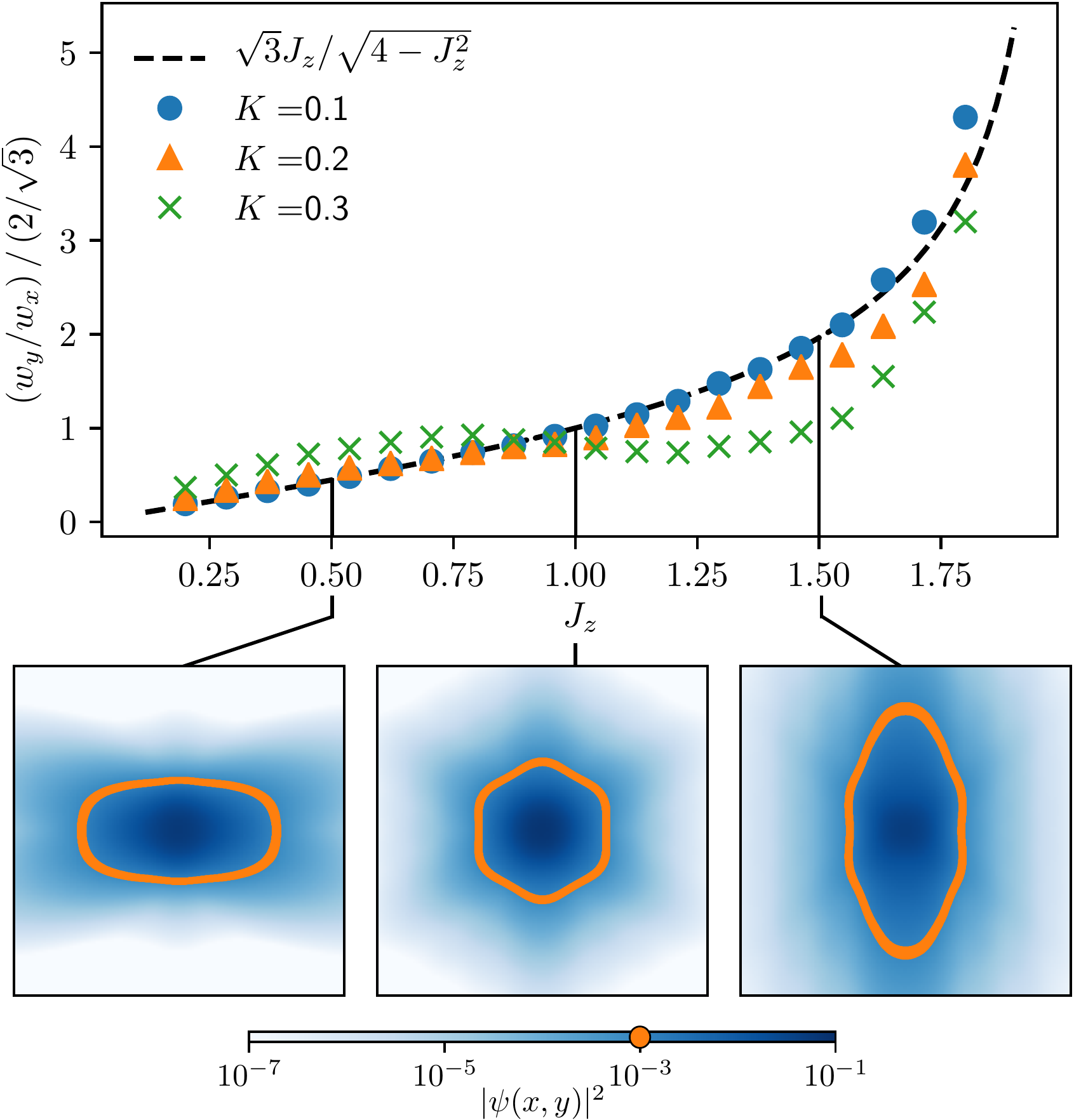}
\caption{Inferring the metric distortions from the spatial profile of a vortex wave function. The points in the main panel plot the ratio between the height and width of the vortex boundary $w_y/w_x$, divided by the height to width ratio of a regular hexagon $2/\sqrt{3}$ for $J_x=J_y=1$, $\epsilon =1$, system size $36 \times 36$ and a range of $K$. Also plotted with a dashed line is the theoretical relationship given in Eq.~(\ref{eq:spaceratio}) we expect from the geometrical description. The numerical data converges to the theoretical line as $K$ decreases. Below are illustrative examples of the hexagonal boundaries we find at various $J_z$ and $K=0.125$. At the isotropic point, $J_z=1$, it is a regular hexagon. As $J_z$ deviates from the isotropic point the ratio $w_y/w_x$ can take smaller or larger values than $2/\sqrt{3}$.}
\label{fig:combined-vortex-data-fig}
\end{figure}

From the continuous profile of the vortex we extract its relative scale in the $x$ and $y$ spatial directions by numerically identifying the set of points where $|\psi(\boldsymbol{r})|^2$ has decayed to a fixed value, e.g. $10^{-3}$. This `boundary' line is drawn for a vortex at the isotropic point of the model in Fig.~\ref{fig:lattice-to-bounding-box}(b). We observe that the geometry of the honeycomb lattice is strongly manifested in this boundary. This is due to the exponential localisation of the zero mode to the vortex, with localisation length comparable to the lattice spacing and independent of the system size if we keep the energy gap constant. Hence, the lattice effects are expected to be visible in the wave function of the zero mode.

As we move away from the isotropic point the hexagon will be stretched in either the $x$ or $y$ direction. To compare the changing shape of the boundary to the distortions predicted by (\ref{eq:spaceratio}), we compare the ratio between the height and width of the hexagonal boundary $w_y/w_x$ to the height and width ratio of a regular hexagon, $(2/\sqrt{3})$. We find that at the isotropic point the boundary satisfies $w_y/w_x=2/\sqrt{3}$. Fig.~\ref{fig:combined-vortex-data-fig} plots the comparison of $(w_y/w_x)/(2/\sqrt{3})$ to the ratio $d_x/d_y$, given in (\ref{eq:spaceratio}), for different values of $J_z$. The ratio $w_y/w_x$ converges to $d_x/d_y$ with decreasing $K$. As $K$ decreases the energy gap also decreases, which leads to an increase in the correlation length. When the correlation length becomes large compared to the lattice spacing discrete lattice effects become negligible and the probability densities approximate the behaviour of those in a continuous system. We find that these ratios agree very well, particularly away from the phase transition boundaries. We observe that an increase (decrease) in $J_z$ equates to a decrease (increase) in the effective distance between lattice sites $d_y$ ($d_x$). This results in the apparent stretching (squeezing) of the zero mode in that direction, i.e. an increase (decrease) in $w_y/w_x$. As a result the effective geometric description of KHLM in terms of a metric is faithful. As the metric is a geometric primitive, we expect that the rest of the geometric quantities, such as the curvature, to be faithfully reproduced as well. We leave this investigation to future work.

 
\section{Conclusions}
\label{sec:Conclusions}

In this paper we expanded upon the known result that the low energy limit, or continuum limit, of the Kitaev honeycomb lattice model is described by massless Majorana fermions obeying the Dirac Hamiltonian embedded in a Minkowski spacetime. We took this idea further by investigating whether the continuum limit could possibly yield non-trivial \textit{curved} geometry. A suitable generalisation of Minkowski spacetime, namely a Riemann-Cartan geometry, with both curvature and torsion manifests itself in the KHLM via non-trivial dreibein $ e_a^{\ \mu} $ and spacetime connection $\Gamma^\rho_{\ \mu \nu} = \tilde{\Gamma}^\rho_{\ \mu \nu} + K^\rho_{\ \mu \nu}$. It was shown that if the couplings of KHLM take a general space-dependent form, a Riemann-Cartan continuum limit can indeed be obtained.

We first showed theoretically that the single-particle Hamiltonian of a Majorana field in a Riemann-Cartan spacetime $h_\text{RC}$ can be identified with the continuum limit of the KHLM Hamiltonian $h_\text{KHLM}$ with general space-dependent nearest-neighbour couplings $\{ J_i \}$ and next-to-nearest-neighbour couplings $\{ K_i  \}$. We demonstrated that the nearest-neighbour part of the Hamiltonian becomes the kinetic term of the Dirac-like continuum limit while the next-to-nearest neighbour part generates an energy gap at the Fermi points. By comparing $h_\text{RC}$ and $h_\text{KHLM}$, we subsequently interpreted the coefficients of the kinetic term as the dreibein corresponding to a non-trivial Levi-Civita connection $\tilde{\Gamma}^\rho_{\ \mu \nu}$ and interpreted the $K$ term that generates the gap as a non-trivial contortion $K^\rho_{\ \mu \nu}$. With this identification the metric $g_{\mu \nu} = e^a_{\ \mu} e^b_{\ \nu} \eta_{a b}$ was fully determined, as well as the total connection $\Gamma^\rho_{\ \mu \nu} = \tilde{\Gamma}^\rho_{\ \mu \nu} + K^\rho_{\ \mu \nu}$, providing us with the curvature $R^\rho_{\ \mu \nu \sigma}$ and torsion $T^\rho_{\ \mu \nu}$ of the Riemann-Cartan theory.

For the special case of isotropic couplings where $J_x = J_y = J_z = J$, we demonstrated that the continuum limit had non-trivial dreibein and spacetime connection yielding a curvature and torsion depending on the parameters $J$ and $K$. The torsion term is equivalent to the superconducting gap. So when the torsion is dominant the system is a topological superconductor in class D. We also demonstrated that by modifying the model with a Kekul\'e distortion, one can give the Majorana fermions a non-zero mass in the continuum limit.  When  the mass term is dominant over the torsion then the system is in class BDI. To study the anisotropic case we considered in detail the special configuration of couplings where $J_x = J_y = 1$, with $J_z$ a free parameter. In this case, the metric of the continuum limit corresponded to a non-uniform dilation in the $x$- and $y$-directions relative to the isotropic metric. This stretching of space was confirmed by analysing the spatial distribution of the quantum correlations as well as the zero-mode profiles of the model. We numerically determined that the description of the model in terms of a geometric metric is faithful even for moderate system sizes. In Fig.~\ref{fig:combined-vortex-data-fig-corr} and Fig.~\ref{fig:combined-vortex-data-fig} we see how well the geometric theory describes the deformation of the quantum correlations and zero mode profiles, respectively. We see the accuracy of the geometric description improves with increasing correlation length. As expected, in the limit of large correlation length discrete lattice effects become negligible and the model strongly resembles its continuous description.

Our work verified that the theory of Majorana fields in Riemann-Cartan geometry can faithfully describe the microscopic Kitaev honeycomb lattice model. This field theoretic description can be then employed to analytically investigate a variety of properties of the microscopic model. As an example, this opens up the exciting possibility to quantitatively study the energy-momentum currents of KHLM and determine their behaviour in terms of various coupling configurations or external driving. The geometric description of superconductors used to obtain the thermal transport coefficients from response theory as employed by Luttinger~\cite{Luttinger} can be realised here with a perturbation of the KHLM couplings around the isotropic and homogeneous configuration. Similarly, our geometric description of KHLM parallels the coupling of topological superconductors to a geometric backgroung~\cite{Palumbo} that demonstrate, with the help of the AdS/CFT correspondence, the presence of chiral edge modes at their boundary.

Our geometric framework can also consider the presence of Heisenberg interactions in the KHLM. In the low energy limit these interactions take the form of four Majorana fermion interactions, $c^a_+ c^a_- c^b_+ c^b_-$. For weak couplings mean field theory can be employed that will renormalise the tunnelling $J$ couplings and the mass $m$ of the Majorana fermions, thus appropriately renormalising the metric of the geometric description. Moreover, chiral gauge fields have been investigated in the context of graphene \cite{Jackiw2} that could be extended to the KHLM case, giving a more complete quantum field theory description of the model. Finally, the response of KHLM to time-dependent geometric perturbations including quenches~\cite{LiuGromovPapic} or considering a dynamical metric introduced by additional quantum fields, such as phonons that perturb the tunnelling couplings, can also be probed using the formalism developed here. We leave these investigations to a future work.

\acknowledgements 

We would like to thank Omri Golan, Duncan Haldane, Paolo Maraner, Jaakko Nissinen, Joost Slingerland, Christopher J. Turner and Julien Vidal for inspiring conversations. This work was supported by the EPSRC grant EP/R020612/1. Statement of compliance with EPSRC policy framework on research data: This publication is theoretical work that does not require supporting research data.

\bibliography{references}


\end{document}